\documentclass[12pt,dvips]{article}
\usepackage[dvips]{graphicx}
\usepackage{graphicx}
\usepackage{latexsym}
\usepackage{cite,./mcite}
\parskip=\itemsep               
\graphicspath{{Plots/}}
\DeclareGraphicsExtensions{.eps,.ps}

\usepackage{amsmath}
\usepackage{latexsym}
\usepackage{psfrag}
\usepackage[english]{babel}

\newcommand{\Tr}{\operatorname{Tr}}
\newcommand{\ad}{\textrm{d}}

\newcommand{\umu}{_{\mu}}
\newcommand{\omu}{^{\mu}}

\newcommand{\umunu}{_{\mu\nu}}

\newcommand{\onu}{^{\nu}}
\newcommand{\unu}{_{\nu}}
\newcommand{\beqn}{\begin{equation}}
\newcommand{\eeqn}{\end{equation}}

\newcommand{\oprime}{^{\,\prime}}

\newcommand{\ii}{\textrm{i}}
\newcommand{\rpeak}{\langle \rho \rangle}
\newcommand{\spf}{\sigma(x)_{\textrm{sp}}}
\newcommand{\rhop}{\langle \rho \rangle}
\newcommand{\muren}{\mu_{\textrm{r}}}

\newcommand{\lwig}{\mbox{\,\raisebox{.3ex}
    {$<$}$\!\!\!\!\!$\raisebox{-.9ex}{$\sim$}\,}}
\newcommand{\gwig}{\mbox{\,\raisebox{.3ex}
    {$>$}$\!\!\!\!\!$\raisebox{-.9ex}{$\sim$}}\,}
\def\Fref#1{Fig.~{\ref{#1}}}
\def\Eref#1{Eq.~({\ref{#1}})}    

\date{\empty}

\title{\vspace{-1cm}{\normalsize\rightline{DESY 08-017}\rightline{UWTHPh-2008-02}}
	\vspace{1cm}
      {\bf QCD-Instantons and Conformal Space-Time Inversion Symmetry}
       \vspace{10mm}} 
\author{D. Klammer\\ 
Fakult\"at f\"ur Physik, Universit\"at Wien,\\ Boltzmanngasse 5, A-1090 Wien, Austria\\[5mm]
and\\[5mm]
F. Schrempp\\
Deutsches Elektronen-Synchrotron DESY, Hamburg, Germany}

\begin{document}
\begin{titlepage} 
  \maketitle
\vspace{1cm}
\begin{abstract}

In this paper, we explore the appealing possibility that the strong suppression of large-size QCD instantons -- as evident from lattice data -- is due to a surviving conformal space-time inversion symmetry. This symmetry is both suggested from the striking invariance of high-quality lattice data for the instanton size distribution under inversion of the instanton size $\rho \to \langle\rho\rangle^2/\rho$ and from the known validity of space-time inversion symmetry in the classical instanton sector. We project the instanton calculus onto the four-dimensional surface of a five-dimensional sphere via conformal stereographic mapping, before investigating conformal inversion. This projection to a compact, curved geometry is both to avoid the occurence of divergences  and to introduce the average instanton size $\rpeak$ from the lattice data as a new length scale. The average instanton size is identified with the radius $b$ of this 5d-sphere and acts as the conformal inversion radius. For $b=\langle\rho\rangle$, our corresponding results are almost perfectly symmetric under space-time inversion and in good qualitative agreement with the lattice data. For $\rho/b\to 0$ we recover the familiar results of instanton perturbation theory in flat 4d-space. Moreover, we illustrate that a (weakly broken) conformal inversion symmetry would have significant consequences for QCD beyond instantons. As a further successful test for inversion symmetry, we present striking implications for another instanton dominated lattice observable, the chirality-flip ratio in the QCD vacuum.
\end{abstract}


\thispagestyle{empty}
\end{titlepage}
\newpage \setcounter{page}{2}

\section{Setting the Stage}
\begin{figure}[b]
\vspace{-3cm}
\includegraphics[scale=0.6,angle=-90]{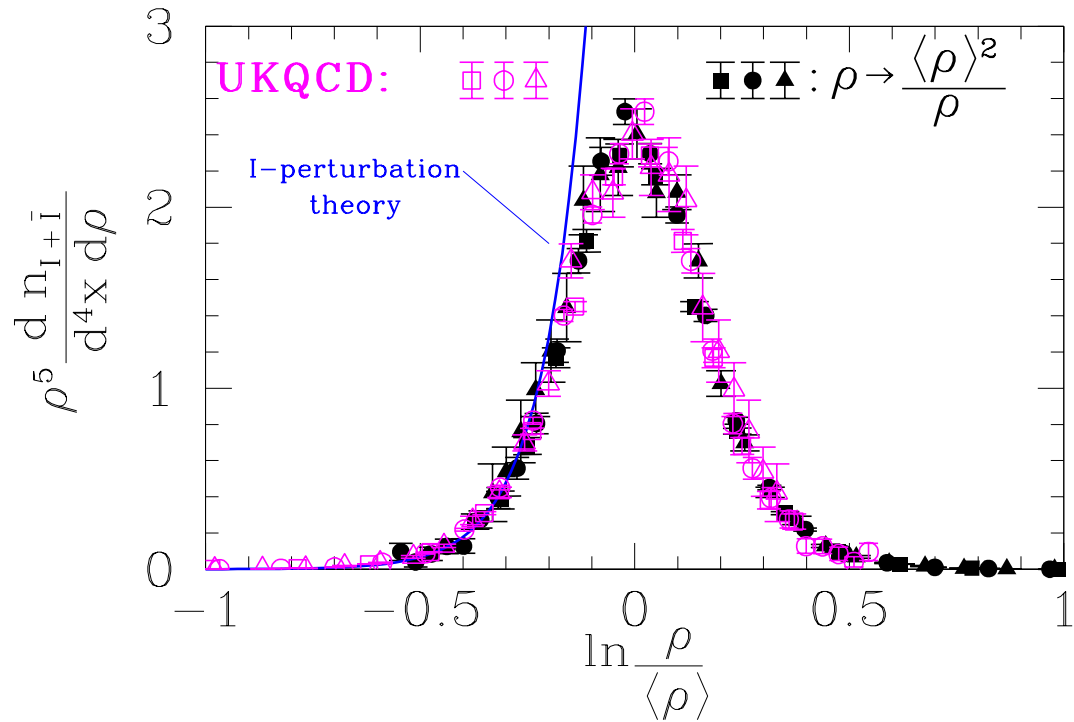}
\caption[dum]{\label{isize} UKQCD lattice data~\cite{Smith:1998wt,Ringwald:1999ze} (open symbols) for the instanton size distribution, displayed such as to suggest a virtually perfect {\it inversion symmetry} under $\rho\rightarrow \rpeak^2/\rho$ with $\rpeak\approx 0.6$ fm
(open and solid data symbols fit onto one universal, symmetric curve). The solid line corresponds to the parameter-free prediction of instanton perturbation theory~\cite{Ringwald:1999ze} using the lattice result $\Lambda_{\overline{\rm MS}\,n_f=0}=(238\pm19)$ MeV from the ALPHA collaboration~\cite{Capitani:1998mq}.}
\end{figure}

Instantons~\cite{Belavin:1975fg,'tHooft:1976fv,'tHooft:1976fv2} represent a basic non-perturbative aspect of Yang-Mills theories in general and QCD in particular. 
One of the most relevant and intriguing quantities within the instanton calculus is the instanton size distribution or instanton density. It has first been derived for small-sized instantons via the vacuum-to-vacuum tunneling amplitude at the one-loop-level of instanton perturbation theory in a seminal paper by 't Hooft~\cite{'tHooft:1976fv,'tHooft:1976fv2}. The instanton size distribution has also been measured in various lattice simulations
~\cite{deForcrand:1997yw,GarciaPerez:1998ru,Smith:1998wt,Ringwald:1999ze,Hasenfratz:1998qk}. Specifically, we shall use throughout this paper the high-statistics data by the UKQCD collaboration~\cite{Smith:1998wt,Ringwald:1999ze}  (cf.~Fig.~\ref{isize} ). 
For instanton sizes $\rho$ smaller than $\sim 0.35~\textrm{fm}$, a parameter-free agreement with instanton perturbation theory has been found~\cite{Ringwald:1999ze}, but a dramatic disagreement appears most rapidly for somewhat larger instantons (cf.~Fig.~\ref{isize} ). In instanton perturbation theory, the weight of larger instantons is growing indefinitely, causing the familiar infrared divergencies of the instanton calculus. Instead, the lattice data exhibit a sharp peak around $\rpeak\approx 0.6$ fm and thereafter, exhibit a strong suppression of large instantons as is also physically expected. Altogether, a satisfactory understanding of the r\^ole of larger-size instantons in the QCD-vacuum is definitely still lacking. It is particularly intriguing that the breakdown of instanton perturbation theory happens so rapidly and dramatically around the appearance of the new length scale $\rhop\cong 0.6 ~\textrm{fm}$, corresponding to the peak position in the lattice data. It is thus clearly worthwhile, to ask what kind of underlying physics could give rise to such a rapid and dramatic change of behaviour of the instanton density.

The purpose of this paper is to discuss and to substantiate an appealing, possible explanation, which was first proposed by one of us~\cite{Schrempp:2001ir}. The central idea is that 
a residual symmetry under conformal inversion of space-time,
\beqn\label{inv_x}
x\umu \rightarrow x\umu\oprime=\frac{\rpeak^2}{x^2}\,x\umu,
\eeqn	 
might be at the root of protecting instantons of becoming too large.

First of all, as apparent from \Fref{isize}, the lattice data~\cite{Smith:1998wt} appear to be invariant under an inversion of the instanton size $\rho$,
\beqn\label{inv}
\rho \Leftrightarrow \rho\oprime=\frac{\rpeak^2}{\rho}.
\eeqn 
The reason for displaying the lattice data in \Fref{isize} versus $\ln(\rho/\rpeak)$ was to make the virtually perfect symmetry under an inversion~(\ref{inv}) of $\rho$
self-evident in the lattice data. Both the open data symbols, referring to
the original data points, and the solid ones, involving inverted
arguments according to \Eref{inv}, fit beautifully  onto one
universal, symmetric curve.   
 
On the theoretical side, the possibility of such an inversion symmetry is particularly appealing, since it may well be a ``relict'' from the known { \it
conformal invariance} of the whole instanton sector at the classical
level\cite{Jackiw:1976dw,Jackiw:1976fs}. 

Let us briefly recall some essential facts about the symmetry under space-time inversion (\ref{inv_x}) at the classical instanton level and why it may be rewritten as an inversion (\ref{inv}) of the instanton size.

Indeed (cf. Appendix A), under a space-time inversion (\ref{inv_x}) the familiar expression for the vector potential of an $SU(2)$ instanton in regular gauge, with gauge coupling $g$, 't Hooft coefficients~\cite{'tHooft:1976fv} $\bar{\eta}_{a\mu\nu}$ and size $\rho$,
\begin{align}
A\umu^{a\,(I)_{\textrm{reg}}}(x;\rho)=\frac{2}{g}\frac{\bar{\eta}_{a\mu\nu}x\onu}{\rho^2+x^2},
\end{align}
 changes into that of an
\emph{anti}-instanton in \emph{singular} gauge of size $\rho\oprime\equiv  \rpeak^2/\rho$~\cite{Jackiw:1976dw},
\begin{align}
\begin{split}\label{invinst}
A\umu^{a\,(I)\,\textrm{reg}}(x;\rho) \rightarrow A\umu^{\prime\,a\,(I)\,\textrm{reg}}(x\oprime;\rho)
=\frac{\partial x\onu}{\partial x^{\prime\mu}}A\unu^{a\,(I)\,\textrm{reg}}(x;\rho)
=A\umu^{a\,(\overline{I})\,\textrm{sing}}(x\oprime;\frac{\rpeak^2}{\rho}).
\end{split}
\end{align}
Using the corresponding conformal transformation law for the field strength tensor under an inversion (\ref{inv_x}),
\begin{align}
G\umunu^{a\, (I)\,\textrm{reg}}(x;\rho) \rightarrow G\umunu^{\prime\, a\,(I)\,\textrm{reg}}(x\oprime;\rho)=
\frac{\partial x^{\alpha}}{\partial x^{\prime \mu}}\frac{\partial x^{\beta}}{\partial x^{\prime \nu}} G_{\alpha\beta}(x;\rho)
=G\umunu^{a\,(\overline{I})\,\textrm{sing}}(x\oprime;\frac{\rpeak^2}{\rho})
\label{invG}
\end{align}
one readily derives for the Lagrange density~\cite{Schrempp:2001ir}
\begin{align}\label{invL}
\begin{split}
\mathcal{L}^{(I)}\left(x,\rho \right)
\rightarrow
\mathcal{L}^{(I)\,\prime}\left(x\oprime,\rho \right)
=\mathcal{L}^{(\overline{I})}\left(x\oprime,\frac{\rpeak^2}{\rho}\right)
\end{split}
\end{align}
The action is of course invariant, since it is independent of the instanton size and the same for instantons and anti-instantons.
\beqn
\int \ad^4 x\oprime\;\mathcal{L}^{(I)\,\prime}\left(x\oprime,\rho \right)=\int \ad^4 x\, \mathcal{L}^{(\bar{I})}\left(x,\frac{\rpeak^2}{\rho} \right)=\int \ad^4 x\;\mathcal{L}^{(I)}\left(x,\rho \right)=\frac{8\pi^2}{g^2}=S_{\textrm{E}}.
\eeqn 
Obviously in Eqs. (\ref{invinst},\ref{invG},\ref{invL}), the coordinate inversion (\ref{inv_x}) has the effect of just \emph{inverting the instanton size}, apart from $I\to\overline{I}$ conjugation and changing the gauge from regular $\leftrightarrow$ singular. An invariance under instanton size inversion for the size distribution is exactly the symmetry indicated by the lattice data (cf. \Fref{isize}). The $I\to\overline{I}$ conjugation is of no concern, since the size distribution as simulated on the lattice, is a sum of both, instantons and anti-instantons. 

The invariance under scale transformations (dilatation) is well-known to be broken at the quantum level via regularisation/renormalisation. While unbroken scale invariance would (nonsensically) make {\it any value} of $\rpeak$ physically equivalent, its breaking signalled by the non-vanishing trace of the energy-momentum tensor, $\theta^\mu_\mu \propto
-\langle0\mid\frac{\alpha_s}{\pi}\,G_{\mu\nu}^{a\,2}\mid 
0\rangle\ne 0$, suggests $\rpeak \sim \langle 0\mid\frac{\alpha_s}{\pi}\,
G_{\mu\nu}^{a\,2}\mid 0\rangle^{-1/4}$. 
   
Let us also recall~\cite{Shintani:1980ve,Shintani:1982vp} in this context the special r\^ole of the space-time inversion $I_{b}$ at some radius $b$. It acts as a discrete conformal transformation that cannot be expressed in infinitesimal form. Hence it cannot be among the 15 generators of the conformal group. Yet the dilatation $\mathcal{D}_\lambda$ and the special conformal transformations $\mathcal{K}_{\alpha_\mu}$ can be expressed by {\em two} inversions $I_a,\ I_b$ of different inversion radii and a translation $\mathcal{T}_{c_\mu}$ (from the Poincar\'e subgroup) by a 4-vector $c_\mu$ ,
\begin{eqnarray}
 \mathcal{D}_{b/a}&=&I_b\,I_{a},\label{dilatation}\\
 \mathcal{K}_{c_\mu/a}&=&I_{a}\mathcal{T}_{c_\mu}I_{a}.
\label{conformal}
\end{eqnarray}
According to \Eref{dilatation}, {\em non-trivial} scale transformations require the space-time inversion symmetry to hold for arbitrary inversion radii. If the inversion radius is instead associated with a {\em physical} scale, the average instanton size $\rpeak \approx 0.6\ \textrm{fm}$, scale transformations $\mathcal{D}_{b/a}$ naturally drop out due to $a=b=\rpeak$ in \Eref{dilatation}, while the inversion $I_{\rpeak}$ may well survive as a symmetry.   

Being defined via the vacuum-to-vacuum tunneling amplitude at the quantum level~\cite{'tHooft:1976fv,CB:qhe}, the full instanton size distribution
represents a difficult challenge with regard to the question of conformal space-time inversion symmetry. Hence, a rigorous proof of the apparent $\rho \to\rpeak^2/\rho$ symmetry is beyond the scope of this investigation. Rather, in this paper, our line of attack is restricted to a detailed study of the {\em zero-mode part} of the size distribution, which we argue constitutes the ``dominating" source of the $\rho$-dependence. Since the zero-modes are closely related to the classical instanton, there is indeed hope that the inversion symmetry is (approximately) preserved. Sect.~\ref{squantum} contains the layout and justification of this underlying strategy. In this context, it is most encouraging that the instanton size distribution of {\em supersymmetric} Yang-Mills theories is known to be entirely given in terms of zero-modes~\cite{Novikov:1983uc}.

Sect.~\ref{sstereo} is central to our approach and also contains our main respective results: We  first project the instanton calculus onto the four-dimensional surface of a five-dimensional sphere via conformal stereographic mapping, before investigating conformal inversion. On the one hand, this projection to a compact, curved geometry avoids the occurence of divergences under space-time inversion. On the other hand, it serves to introduce the average instanton size $\rpeak$ from the lattice data as a crucial length scale through its identification with the radius $b$ of this 5d-sphere, acting as the conformal inversion radius. 
   
In Sect.~\ref{sflip}, we shall briefly discuss some direct, alternative evidence for space-time inversion from the lattice data for a completely independent (lattice) observable, the chirality-flip ratio $R^{\mathrm{NS}}$ in the QCD vacuum~\cite{Faccioli:2002xf,Faccioli:2003qz}.

The validity of our proposed inversion symmetry would allow to access the non-perturbative regime of large-size instantons (yet with {\it small} $\rho\oprime=\rpeak^2/\rho$) in terms of instanton perturbation theory for instantons with {\it small} $\rho\oprime$. It may well have important consequences beyond instanton physics for QCD in general. This intriguing possibility will be addressed towards the end, in Sect.~\ref{sbeyond}.   

\section{Inversion Symmetry at the Quantum Level?\label{squantum}}

Since the conformal space-time inversion symmetry connects the physics at short and long distances, it appears very interesting to investigate its possible validity beyond the classical instanton level in more rigorous terms. This section and the following one are devoted to this non-trivial task.

Let us start from the vacuum-to-vacuum transition amplitude at one-loop level~\cite{'tHooft:1976fv,CB:qhe} which directly
determines the leading expression for the instanton size distribution $d\left(\rho\muren,\, \alpha_s(\muren)\right)$,
\begin{align} \label{vacdens}
\langle 0 \vert 0 \rangle^{(I)}
=\int \prod_i \ad\gamma_i J\left(\gamma\right)Q\left(\gamma\right)\textrm{e}^{-\frac{2\pi}{\alpha_s}}
=\int \ad^4z\frac{\ad\rho}{\rho^5}\, d\left(\rho\muren,\,\alpha_s(\muren)\right)
\end{align}
with the integrations on the left of \Eref{vacdens} extending over all collective coordinates $\gamma_i\in \gamma=\{U,\,z_\mu,\,\rho\}$,  
i.e. the colour orientation matrices $U$, the position $z\umu$ and the size $\rho$ of the instanton.

As detailed by Bernard~\cite{CB:qhe} and apparent in \Eref{vacdens}, the size distribution factors into contributions from zero modes and non-zero modes as follows.
 
The quantity $J\left(\gamma\right)$ is the collective-coordinate Jacobian and thus originates from the various zero modes $\psi^{(i)}$ as,
\begin{align}\label{zm1}
J\left(\gamma\right)&=\left( \prod_i \frac{1}{\sqrt{2\pi}}\right) \left(\det\,\mathcal{U} \right)^{1/2}
=\left( \prod_i \frac{1}{\sqrt{2\pi}}\right) \|\psi^{(i)}\| ^{1/2}
\end{align}
due to the orthogonality of the zero modes with normalisations
\begin{align}\label{zm2}
\begin{split}
\mathcal{U}_{\,ij}=2\,\int\ad^4 x \Tr\left[\psi\umu^{(i)}(x)\,\psi^{(j)\mu}(x) \right]
&=\int\ad^4 x\, \psi^{(i)}_{\mu a}\,\psi^{(j)\mu a}=\delta_{ij}\,\|\psi^{(i)}\|^2 \quad
\textrm{where} \quad
\psi\umu^{(i)}\equiv \psi_{\mu a}^{(i)}\,\frac{\lambda_a}{2},\\
\|\psi^{(i)}\|^2&=2\int\ad^4 x \Tr\left[\psi\umu^{(i)}(x)\,\psi^{(i)\mu}(x) \right].
\end{split}
\end{align}
The quantity $Q\left(\gamma\right)$ in \Eref{vacdens} contains the remaining determinants from Gaussian functional integration taken over the non-zero mode parts as indicated by the primes.  
\begin{align}
Q\left(\gamma\right)\equiv\frac{[\det^{-1/2}\,M_\mathrm{A}\oprime(\gamma)\,\det M_\mathrm{gh}(\gamma)]_{A^{cl}=A^{(I)}}}{[\det^{-1/2} M_{A}\oprime\,\det\,M_\mathrm{gh} ]_{A^{cl}=0}}.
\end{align}

The resulting dimensionless size distribution $d\left(\rho\muren,\,\alpha_s(\muren)\right)$, as introduced on the right of \Eref{vacdens}, is known~\cite{'tHooft:1976fv,'tHooft:1976fv2,Morris:1984zi}  to take the following form for sufficiently small instanton sizes $\rho$,  
\begin{align}\label{density}
d\left(\rho\muren,\,\alpha_s(\muren)\right)=\rho^5\,\frac{\ad n^{(I)}}{\ad^4 z \,\ad\rho}
=C \left(\frac{2\pi}{\alpha_s(\muren)}\right)^{2N_c}\exp\left(-\frac{2\pi}{\alpha_s(\muren)}\right)\left(\rho\muren\right)^b,
\end{align}
with known, scheme dependent constant $C$, renormalisation constant $\muren$ and exhibiting renormalisation group invariance at  
\begin{align}
\left.\begin{array}{l}
\mathrm{1-loop\ level}\\
\mathrm{2-loop\ level}\end{array}\right\}\mathrm{\ for\ }
b=\left\{\begin{array}{lllcl}
\beta_0;&\alpha_s(\muren)&=&\alpha_s^\textrm{1-loop}(\muren)\\
\beta_0+(\beta_1-4 N_c\,\beta_0)\,\frac{\alpha_s(\muren)}{4\pi};&\alpha_s(\muren)&=&\alpha_s^\textrm{2-loop}(\muren).
\end{array}\right.
\end{align}
in terms of the first two coefficients of the QCD $\beta$-function
\begin{align}
\beta_0=\frac{11}{3}N_c-\frac{2}{3}n_f;\quad \beta_1=\frac{34}{3}N_c^2-\left(\frac{13}{3}N_c-\frac{1}{N_c}\right)n_f.
\end{align}
The resulting $\rho$ dependence at two-loop level is displayed in \Fref{isize}.

We are now ready to present the strategy we are going to pursue.

First of all, we know that the classical instanton gauge field manifestly reacts to a space-time inversion with an instanton size inversion, $\rho\to\rho\oprime=\rhop^2/\rho$. Secondly, zero modes are just derivatives of the classical gauge field with respect to the collective coordinates $\gamma_i$ (apart from possible gauge transformations), 
\begin{align}
\psi^{(i)}\umu(x)\sim \frac{\partial A^{(I)}\umu(x;\gamma)}{\partial\gamma_i}.
\end{align}
Hence there are good reasons to hope that the size inversion symmetry is inherited by the entire zero mode contribution to the instanton size distribution. This crucial part of the task will be explicitly studied further below. The result will formally not be restricted to small values of $\rho$.
 
Thirdly, the instanton size distribution appears to be dominated by the zero mode (ZM) contribution $J(\gamma)$ to exponential accuracy, since $4N_c \gg N_c/3$ and since from \Eref{density}
\begin{align}\label{eq:approximation}
\begin{split}
d(\rho\muren,\alpha_s)_{1-\textrm{loop}}&\propto(\rho)^{\beta_o}=(\rho)^{4\,N_c-\frac{N_c}{3}}=\underbrace{(\rho)^{4\,N_c}}_{\textrm{\small{ZM part $\rho^5\,J(\gamma)$}}}\cdot
\underbrace{(\rho)^{-\frac{1}{3}N_c}}_{\textrm{\small{non-ZM part $Q(\gamma)$}}}.
\end{split}
\end{align}
In supersymmetric Yang-Mills theory, the zero mode part becomes even more important\footnote{We are grateful to Mikhail Shifman for pointing this out to one of us (F.S.).}. In this case, the size-distribution $d^{\,\textrm{SUSY}}(\rho\muren,\alpha_s(\muren))$ is entirely determined by $J(\gamma)$, as was shown in~\cite{Novikov:1983uc}, since all non-zero mode contributions cancel precisely to any order in perturbation theory,
see also~\cite{Amati:1988ft,Novikov:1983uc}. 

Within the regime of instanton perturbation theory, the non-zero mode contribution was first shown by 't Hooft~\cite{'tHooft:1976fv} to yield 
\begin{align}
Q(\gamma)\propto \muren^{4\,N_c}\,\left(\frac{1}{\rho\muren}\right)^{\frac{N_c}{3}}\stackrel{SU(3)}=\muren^{4\,N_c}\,\frac{1}{\rho\mu_{\textrm{r}}},
\label{nzm}
\end{align} 
Since $Q(\gamma)$ becomes infrared sensitive for large $\rho$, rigorous results about the effects of space-time inversion in $Q(\gamma)$ are beyond the scope of this paper. Hence, taking recourse to the argued dominance of the zero mode contribution, we shall heuristically carry along $Q(\gamma)\propto\rho^{-Nc/3} = 1/\rho$ from \Eref{nzm} as a ``correction" factor for all values of $\rho$. This leads us to investigate the $\rho\to\rho\oprime=\rhop^2/\rho$ symmetry for the following approximate form of the instanton size distribution   
\begin{align}\label{dapprox}
d\left(\rho\muren,\,\alpha_s(\muren)\right)\propto \rho^5\,J\left(\rho\right)\frac{\muren^{4\,N_c}}{(\muren\rho)^{Nc/3}}\exp\left\{- \frac{2\pi}{\alpha_s{(\mu_{\textrm{r}})}}\right\},
\end{align}   
with the zero mode contribution $J(\rho)$ given via Eqs.~(\ref{zm1}, \ref{zm2}).  

Specializing next to $SU(3)$ instantons, we encounter four types of zero modes: one dilatation zero mode $\psi^{(\rho)}\umu(x)$, four translation zero modes $\psi^{(z)}\umu(x)$ and two types of colour zero modes\footnote{This is because the generators $\lambda_i$, $i=1\hdots7$, of
the gauge group $SU(3)$ are grouped into different multiplets (a triplet and two doublets) with respect to the $SU(2)$ subgroup.}: three colour zero modes coming from the generators $\lambda_1, \lambda_2, \lambda_3$, $\psi^{(a)}\umu(x)$, and finally four colour modes coming from the remaining
generators $\lambda_4,...,\lambda_7$, denoted as $\psi^{(\alpha)}\umu(x)$. All in all we have 12 collective coordinates parameterising the instanton configuration for an $SU(3)$ gauge group. The corresponding twelve zero modes are listed in Appendix B. Their normalisations~\cite{CB:qhe} are given by
\begin{equation}\label{zm0}
\frac{1}{\sqrt{2}}\|\psi^{(\rho)}\|=\|\psi^{(z)}\|
=\frac{1}{\rho\sqrt{2}}\|\psi^{(a)}\|=\frac{1}{\rho}\|\psi^{(\alpha)}\|
=\sqrt{\frac{2\pi}{\alpha_s}},
\end{equation}
yielding a total zero mode contribution
\begin{align}\label{eq:zm_contribution}
\begin{split}
\rho^5\,J\left(\gamma\right)
&=
\rho^5\,\frac{
\|\psi^{(\rho)}\|\,\|\psi^{(z)}\|^4
\,\|\psi^{(a)}\|^3\,\|\psi^{(\alpha)}\|^4
}{\left(2\pi\right)^6}\\
&=\frac{1}{2^4\pi^6}\left(\frac{2\pi}{\alpha_s}\right)^6\,\rho^{12}.
\end{split}
\end{align}

In order to study the behaviour of the zero mode part under conformal inversion, we have to apply the transformations of Eq. (\ref{eq:coVF_law}) and Eq.(\ref{eq:contraVF_law}) in Appendix A to the normalisation integrals~(\ref{zm2}). But we encounter number of obvious deficiencies from the start that require a basically modified procedure.
\begin{itemize} 
\item[i)] While the normalisation integrals of the \emph{inverted} dilatation colour zero modes, $\psi^{\,(\rho)\,\prime}\umu(x\oprime,\rho)$ and $\psi^{\,(a)\,\prime}\umu(x\oprime,\rho)$, give a finite result, the normalisations for the inverted translation zero modes $\psi^{\,(z)\,\prime}\umu(x\oprime,\rho)$ 
and the inverted colour zero modes $\psi^{\,(\alpha)\,\prime}\umu(x\oprime,\rho)$ turn out to be \textit{divergent}. Hence a suitable regularization procedure is obviously required before any further statements can be made.
\item[ii)] After application of a space-time inversion transformation to the convergent dilatation and colour mode normalisation integrals, the $\rho$ dependence is indeed modified, yet the normalisation integrals turn out not to be invariant under space-time inversion. For example,
\begin{align}
\Arrowvert \psi^{\,(a)}(\rho) \Arrowvert\Rightarrow\Arrowvert \psi^{\,(a)\,\prime}(\rho) \Arrowvert&=\sqrt{\frac{4\pi}{\alpha_s}}\rho\oprime=\Arrowvert \psi^{\,(a)}(\rho\oprime) \Arrowvert\ne \Arrowvert \psi^{\,(a)}(\rho) \Arrowvert.\label{norminva}
\end{align}
\item[iii)] From the lattice data it is apparent that we first have to incorporate the conspicuous instanton scale $\rpeak$ into the framework, before we can hope for a satisfactory peak description of the instanton size distribution with a symmetry under $\rho\to\rpeak^2/\rho$. The simple monomials in $\rho$ from leading order of instanton perturbation theory~(\ref{zm0}) are  certainly inadequate.
\end{itemize}
The next section is devoted to an elegant resolution of these difficulties and requirements i)-iii).

\section{Implementing the Instanton Scale $\rpeak$\label{sstereo}}

We achieve the finiteness and invariance of all zero mode normalisation integrals under conformal space-time inversion and the introduction of the desired new scale $\rhop$ into the instanton calculus by projecting the 4-dimensional Euclidean
space onto the surface of a sphere, embedded in 5-dimensional Euclidean space~\cite{Adler:1972fx,Adler:1973ty} via stereographic
projection\footnote{From now on we will use Latin indices for the
5-dimensional space whereas the Greek indices run as usual from $1\hdots 4$. Hatted quantities were subject to a stereographic mapping, the prime denotes as before an inversion in Euclidean space-time.},
\begin{align}\label{eq:sp}
x\umu \rightarrow r_a&=\left(r_{\nu},r_5\right),
\end{align}
where
\begin{align}\label{sp}
\begin{split}
r_{\nu}&=\frac{2\, \rpeak^2 \, x_{\nu}}{\rpeak^2+x^2},
\\
r_5&=\rpeak\,\frac{\rpeak^2-x^2}{\rpeak^2+x^2},
\end{split}
\end{align}
such that 
\begin{align}
r^2=r_ar^a=r\unu r\onu + r_5^2=\rpeak^2.
\end{align}
\begin{figure}[t]
\begin{center}
\psfrag{Q}{\textsf{Q}}
\psfrag{Qprime}{$\textsf{Q}\oprime$}
\psfrag{P}{\textsf{P}}
\psfrag{Pprime}{$\textsf{P}\oprime$}
\psfrag{N}{\textsf{N}}
\psfrag{S}{\textsf{S}}
\psfrag{rfive}{$r_5$}
\psfrag{rmu}{$r\umu$}
\psfrag{rfive2}{$r_5$}
\psfrag{rfive3}{-$r_5$}
\psfrag{b}{\hspace{-2ex}$\langle\rho\rangle$}
\includegraphics[scale=0.7]{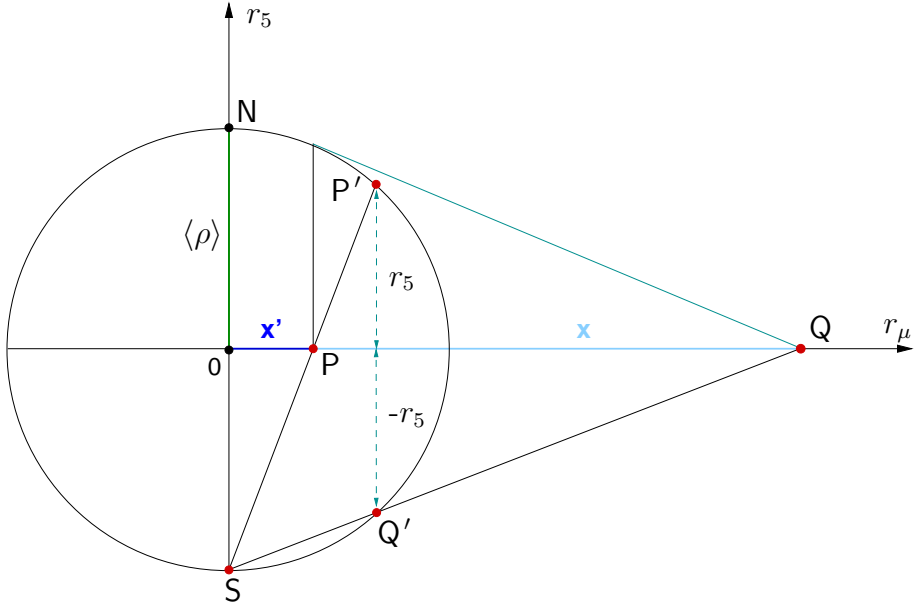}
\caption{\small{Relation between inversion and stereographic projection: The points \textsf{P} and \textsf{Q} define the two distances $x$ and
$x\oprime$, which are related by the condition of
inversion, $x^2\,x^{\,\prime\,2}=\rpeak^4$. As a consequence, the projected points $\textsf{P}\oprime$ and $\textsf{Q}\oprime$ differ only in the sign of the coordinate
$r_5$. An inversion leads to an exchange of the northern and southern hemisphere.}  }
\label{fig:stereo_inversion}
\end{center}
\end{figure}
This mapping was done for the first time by Jackiw and Rebbi~\cite{Jackiw:1976dw}, where an $O(5)$-covariant instanton calculus was developed. However, this was only possible by taking the radius of the spere to be equal to the size parameter $\rho$ of the instanton.  The crucial difference in our approach lies precisely in this radius: In our case it meant to  have a \emph{fixed}, physical value. We \emph{identify the radius of the sphere with the average size of an instanton}: $r^2=\rpeak^2 \approx (0.6\textrm{\ fm})^2$. Of course, we shall also study the interesting limits 
$\rpeak\to0,\ \infty$ further below, in order to reconnect with the known results of instanton perturbation theory.

The sphere, being a compact and curved manifold, will at the same time serve as regulator for the divergent normalisation integrals. The stereographic projection is a conformal transformation itself, corresponding to a scale factor 
\begin{align}
\sigma_{\textrm{sp}}(x,\rpeak)=\frac{\left(\rpeak^2+x^2\right)^2}{4\rpeak^4}.
\end{align}
The relation between the $4$-dimensional Euclidean volume element and the area element on the hypersphere turns out to be quite remarkable:
\begin{align}\label{eq:b}
\begin{split}
\ad  A &=\frac{1}{\sigma_{\textrm{sp}}\,(x,\rpeak)^2}\ad^4x\\
&= \frac{16\, \rpeak^8}{(\rpeak^2+x^2)^4}\ad^4 x.
\end{split}
\end{align}
We find that -- apart from a multiplicative constant -- the conformal factor $\sigma_{sp}^{-2}(x,\rho)$ just equals the Lagrange density of an instanton in 4-dimensional flat space, 
\beqn 
\mathcal{L}^{(I)}=\frac{2}{\pi\alpha_s}\frac{6\,\rho^4}{(\rho^2+x^2)^4},
\eeqn 
if the radius of the sphere corresponds to the instanton size $\rho$. This somewhat surprising relation is due to the functional form of the instanton field and suggests that the 5-dimensional sphere is a natural environment for studying symmetry properties of instanton physics that are eventually hidden in  Euclidean space.

The projection of the instanton calculus onto the sphere is a straight forward application of the rules of Eq.(\ref{eq:coVF_law}) and Eq.(\ref{eq:contraVF_law}). The only difference being that here we go from  four space-time coordinates to five space-time coordinates plus one constraint, which requires some care when inverse transformations are involved (see Appendix A).  

Let us also note that the entire set of the four translation zero modes in background gauge, has the form of the field strength tensor~\cite{CB:qhe},
\begin{align}
\psi\umu^{(z_\nu)}\left(x\right)=F_{\mu\nu}(x).                               \end{align}
Nevertheless we are interested in the lengths of the four individual translational zero modes, each representing a vector field. 
Thus only one space-time index of the collectivity of translation zero modes is affected by a  conformal inversion (cf. Appendix A),
\begin{align}\label{eq:transZMafterInv}
\psi^{\,(z_\nu)\,\prime}\umu(x\oprime)&=\sqrt{\sigma_{\textrm{inv}}(x)}I^{\phantom{i}\sigma}\umu(x)\,\psi^{(z_\nu)}_{\sigma}(x);\quad \nu=1\ldots 4.
\end{align}
This puts the behaviour of normalisation integrals under inversion\footnote{This applies actually to all conformal transformations, notably to stereographic projections.}  for all types of zero modes on the same footing,
and is in accordance with the computation in~\cite{CB:qhe}, where the total contribution in Eq.(\ref{eq:zm_contribution}) is the product of the indiviual twelve zero modes. 

Since the stereographic projection and the inversion are both conformal transformations,
it is not surprising that the conformal factor $\sigma_{\mathrm{sp}}(x,\rpeak)$ transforms under inversion as  
\begin{align}\label{eq: sp_inv}
\sigma_{\mathrm{sp}}(x\oprime,\rpeak)
=\frac{\sigma_{\mathrm{sp}}(x,\rpeak)}{\sigma_{\mathrm{inv}}(x,\rpeak)}.
\end{align}
Due to \Eref{eq: sp_inv}, the area element~(\ref{eq:b}) on the 5-dimensional sphere transforms under conformal inversion in Euclidean space as
\begin{align}\label{Ainv}
\ad A\oprime = \frac{\ad^4 x\oprime} {\sigma_{\textrm{sp}}^2(x\oprime,\rpeak)}
=\frac{\ad^4 x}{\sigma_{\textrm{inv}}^2(x,\rpeak)\,\sigma_{\textrm{sp}}^{2}(x\oprime,\rpeak)} =\frac{1}{\sigma_{\textrm{sp}}\,(x,\rpeak)^2}\ad^4x =\ad A
\end{align}
and is thus invariant.

Let us stress that this result holds since according to our approach, the radius of the 5-dimensional sphere was identified with the radius of inversion\footnote{Relations for different radii of sphere and inversion can be found in~\cite{Klammer:2006cq}.}, i.e. the average instanton size $\langle \rho \rangle$.

We lift the zero modes, which we have computed in Euclidean space first, by means of stereographic projection to the sphere. For the normalisation integrals of the zero modes projected on the sphere we have to evaluate the
following expression applying Eq.(\ref{eq: stp trafo rules}),
\begin{align}\label{eq:ZM_Norm_ints_sphere}
\begin{split}
\| \widehat{\psi}\left(\rho\right)\|^2
&= 2\
\int \ad  A \,\Tr\left[\widehat{\psi}_a(r)\widehat{\psi}^a(r)\right]
\\
&=2\,\int \ad^4 x\,\sigma_{\mathrm{sp}}^{-1}(x,\rpeak)\Tr\left[\psi\umu(x)\psi\omu(x)\right],
\end{split}
\end{align}
and  compare now Eq. (\ref{eq:ZM_Norm_ints_sphere}) to the normalisation integral of the inverted zero modes $\widehat{\psi}^{\,\prime}$
projected onto a sphere with radius $\rhop$. 
We find readily
\begin{align}\label{eq:ZM_on_sphere_inv}
\| \widehat{\psi}\oprime\left(\rho\right)\|^2
=2\,\int \ad  A\oprime  \,\Tr\left[\widehat{\psi}^{\,\prime}_a(r^\prime)\widehat{\psi}^{a\,\prime\,}(r^\prime)\right]=2\
\int \ad  A \,\Tr\left[\widehat{\psi}_a(r)\widehat{\psi}^a(r)\right]=\| \widehat{\psi}\left(\rho\right)\|^2
\end{align}
due to $\ad A\oprime=\ad A$ from \Eref{Ainv} and with the help of Eqs.~(\ref{eq: stp trafo rules}, \ref{eq: sp_inv}), 
\begin{align}
\Tr\left[\widehat{\psi}^{\,\prime}_a(r^\prime)\widehat{\psi}^{a\,\prime\,}(r^\prime)\right] 
&=\sigma_{\mathrm{sp}}(x\oprime,\rpeak)\,\Tr\left[\psi\oprime\umu\left(x\oprime\right)\psi^{\,\mu\,\prime}\left(x\oprime\right)\right]
=\sigma_{\mathrm{sp}}(x\oprime,\rpeak)\,\sigma_{\mathrm{inv}}(x,\rpeak)\;\Tr\left[\psi\umu(x)\psi\omu(x)\right]\nonumber\\
&=\sigma_{\mathrm{sp}}(x,\rpeak)\,\Tr\left[\psi\umu(x)\psi\omu(x)\right]=\Tr\left[\widehat{\psi}_a(r)\widehat{\psi}^a(r)\right].
\end{align}
Since the radius of the sphere and the radius of inversion are equal, 
we find that \textit{all normalisation integrals of the zero modes are invariant under conformal inversion.} This encouraging fact is a central result of our approach. It notably implies a particular form of the resulting $\rho$ dependence of the zero mode contribution, which will be examined next. Another important implication is that the troublesome translation and colour zero modes, $\psi^{(z)}(x)$ and $\psi^{(\alpha)}(x)$, now retain \emph{finite} normalisation integrals under space-time inversion. 

Next, let us turn to the crucial question: Does the invariance~(\ref{eq:ZM_on_sphere_inv}) of the total zero mode contribution 
\begin{align}\label{allzms}
 \widehat{J}(\kappa) =\frac{1}{\left(2\pi\right)^6}\Arrowvert\widehat{\psi}^{(\rho)}\Arrowvert
 \times
 \Arrowvert\widehat{\psi}^{(z)}\Arrowvert^4
 \times
 \Arrowvert\widehat{\psi}^{(a)}\Arrowvert^3
 \times
 \Arrowvert\widehat{\psi}^{(\alpha)}\Arrowvert^4
\end{align}
under space-time inversion reflect in an 
instanton size distribution~(\ref{dapprox}) [with $\muren\approx 1/\rpeak$ ] that is (approximately) invariant under inversion of the instanton size?

To this end, we evaluate first the various types of zero mode norms individually, introducing the dimensionless variable, 
\begin{equation}\label{kappa}
\kappa=\frac{\rho}{\rpeak},
\end{equation}
such that
\begin{align}
\rho \rightarrow \rho\oprime=\frac{\rhop^2}{\rho}
\quad \Leftrightarrow \quad \kappa \rightarrow \frac{1}{\kappa}
\quad \Leftrightarrow \quad \ln(\kappa) \rightarrow -\ln(\kappa).
\end{align}

We find the following results on the sphere:
\paragraph*{Dilatation zero mode:}

\begin{align}
\begin{split}
\| \widehat{\psi}^{(\rho)}(\kappa)\|^2
&=
2\,\int\ad^4 x \sigma_{\textrm{sp}}^{-1}(x,\rpeak)\,\Tr\left[\psi^{(\rho)}\umu(x)\,\psi^{\,\mu\,(\rho)}(x)\right]
=
\frac{1}{\pi\alpha_s}\int\ad^4 x \frac{48\rpeak^4}{\left(\rpeak^2+x^2\right)^2}\frac{\rho^2 x^2}{\left(\rho^2+x^2\right)^4}
\\
&=\frac{16\pi}{\alpha_s}\left(
-\frac{12\kappa^2\left(1+\kappa^2\right)\ln\left(\kappa\right)}{\left(\kappa^2-1\right)^5}
+ \frac{\left(\kappa^4+10\kappa^2+1\right)}{\left(\kappa^2-1\right)^4}\right)
=\frac{1}{\kappa^4}\,\| \widehat{\psi}^{(\rho)}(\frac{1}{\kappa})\|^2
\end{split}
\end{align}
The functional form of this normalisation integral is preserved up to a scaling factor for $\kappa$.
Therefore we find that 
\begin{align}
\kappa\,\| \widehat{\psi}^{(\rho)}(\kappa)\| = 
\frac{1}{\kappa}\,\| \widehat{\psi}^{(\rho)}(\frac{1}{\kappa})\|
\end{align}
is symmetric under inversion of $\kappa\Leftrightarrow 1/\kappa$, while 
$\| \widehat{\psi}^{(\rho)}(\kappa)\|$ is not.

\paragraph*{Colour zero modes for $\lambda_a;\ a=1,\,2,\,3$:}
\begin{align}
\begin{split}
\| \widehat{\psi}^{(a)}(\kappa)\|^2
&=2\,\int\ad^4 x \sigma_{\textrm{sp}}^{-1}(x,\rpeak)\,\Tr\left[\psi^{(a)}\umu(x)\,\psi^{\,\mu\,(a)}(x)\right]
=
\frac{1}{\pi\alpha_s}\int\ad^4 x \frac{48\rpeak^4}{\left(\rpeak^2+x^2\right)^2}\frac{\rho^4 x^2}{\left(\rho^2+x^2\right)^4}
\\
&=
\frac{16\pi}{\alpha_s}\rpeak^2\kappa^2\, \left(
-\frac{12\kappa^2\left(1+\kappa^2\right)\ln(\kappa)}{\left(\kappa^2-1\right)^5}
+\frac{\left(\kappa^4+10\kappa^2+1\right)}{\left(\kappa^2-1\right)^4}\right)
=
\|\widehat{\psi}^{(a)}(\frac{1}{\kappa})\|^2
\end{split}
\end{align}
The colour zero modes are exactly symmetric under $\kappa\Leftrightarrow 1/\kappa$.
Moreover, as in flat 4-dimensional space, \Eref{zm0}, they satisfy
\begin{align}
\| \widehat{\psi}^{(a)}(\rho)\|=\rho\,\| \widehat{\psi}^{(\rho)}(\rho)\|
\end{align}

\paragraph*{Colour zero modes for $\lambda_{\alpha};\ \alpha=4\ldots 7$:}
\begin{align}
\begin{split}
\| \widehat{\psi}^{(\alpha)}(\kappa)\|^2
&=2\,\int\ad^4 x \sigma_{\textrm{sp}}^{-1}(x,\rpeak)\,\Tr\left[\psi^{(\alpha)}\umu(x)\,\psi^{\,\mu\,(\alpha)}(x)\right]
=
\frac{1}{\pi\alpha_s}\int\ad^4 x \frac{16\rpeak^4}{\left(\rpeak^2+x^2\right)^2}\frac{1}{\left(\rho^2+x^2\right)^3}
\\
&=
\frac{8\pi}{\alpha_s}\,\rhop^2\kappa^2\, \left[
\frac{4\kappa^2\left(2+\kappa^2\right)\ln(\kappa)}{\left(\kappa^2-1\right)^4}
-\frac{\left(1+5\kappa^2\right)}{\left(\kappa^2-1\right)^3}\right]
\end{split}
\end{align}
For these colour zero modes, the functional form of the normalisation integral changes under an inversion of the instanton size and thus it is not  symmetric. 

\paragraph*{Translation zero modes:}
\begin{align}
\begin{split}
\| \widehat{\psi}^{(z)}(\kappa)\|^2
&=2\,\int\ad^4 x \sigma_{\textrm{sp}}^{-1}(x,\rpeak)\,\Tr\left[\psi^{(z)}\umu(x)\,\psi^{\,\mu\,(z)}(x)\right]
=
\frac{1}{\pi\alpha_s}\int\ad^4 x \frac{48\rpeak^4}{\left(\rpeak^2+x^2\right)^2}\frac{\rho^4 }{\left(\rho^2+x^2\right)^4}\\
&=
\frac{8\pi}{\alpha_s}\left(\frac{12\kappa^4\left(\kappa^2+3\right)\ln(\kappa)}{\left(\kappa^2-1\right)^5}
-\frac{\left(17\kappa^4+8\kappa^2-1\right)}{\left(\kappa^2-1\right)^4}\right)
=
\|\widehat{\psi}^{(z)}(\frac{1}{\kappa})\sqrt{\sigma_{\textrm{inv}}}\|^2
\end{split}
\end{align}
The functional form of this normalisation integral changes under an inversion of $\rho$, thus this zero mode is not symmetric under the desired transformation. 

\begin{figure}
\begin{center}
\parbox{6cm}{
\psfrag{logarithmus}{$\ln \left(\kappa\right)$}
\psfrag{baum}{$\| \widehat{\psi}^{(\rho)}\|$}
\includegraphics[scale=0.25,angle=270]{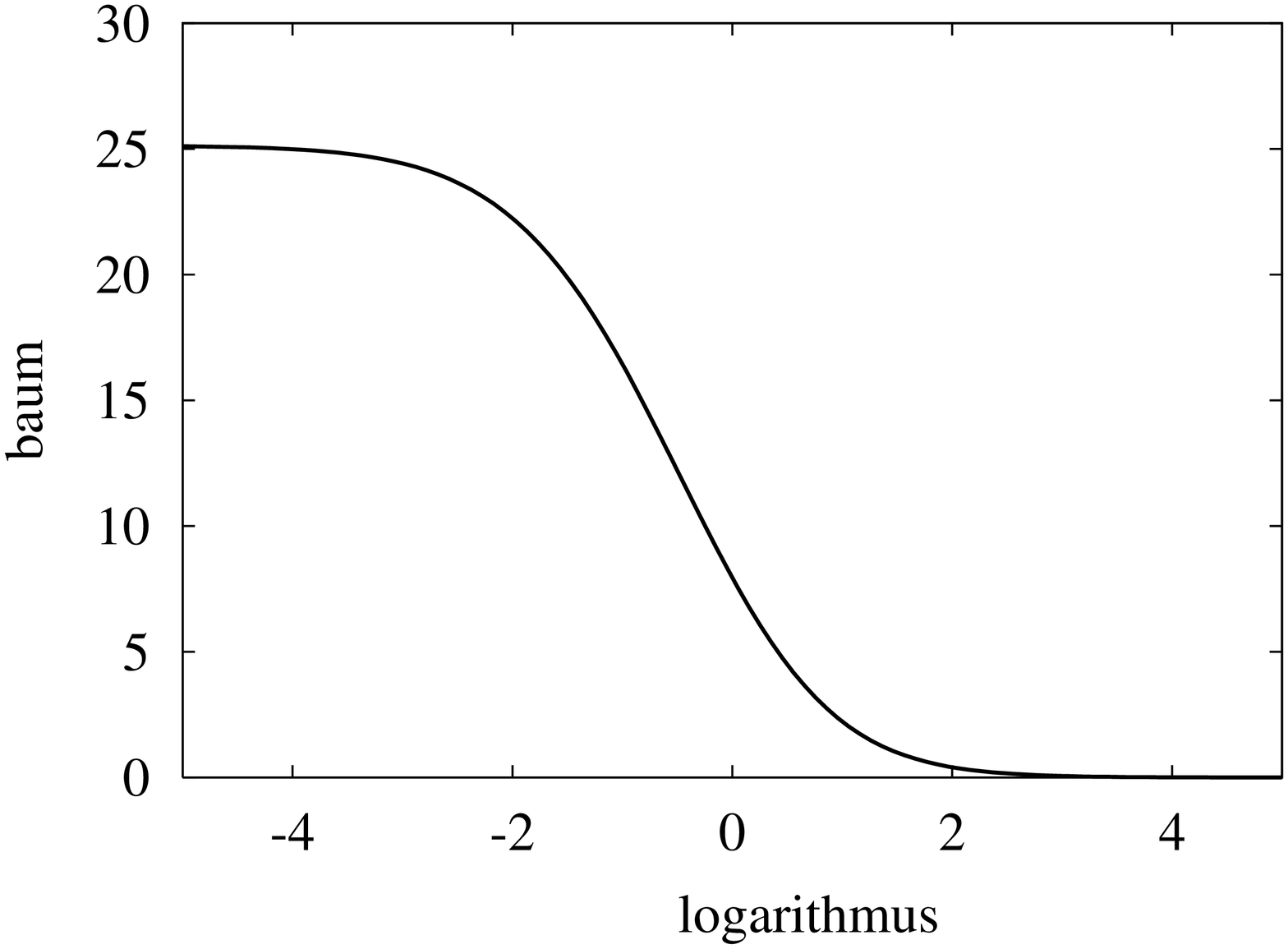}
\caption{\small{Dilatation zero mode}}
\label{fig:o5_dil_ZM}
}
\hspace{2cm}
\parbox{6cm}{
\psfrag{logarithmus}{$\ln \left(\kappa\right)$}
\psfrag{biene}{\hspace{-2ex}$\| \widehat{\psi}^{(a)}\|\big/\rpeak$}
\includegraphics[scale=0.25,angle=270]{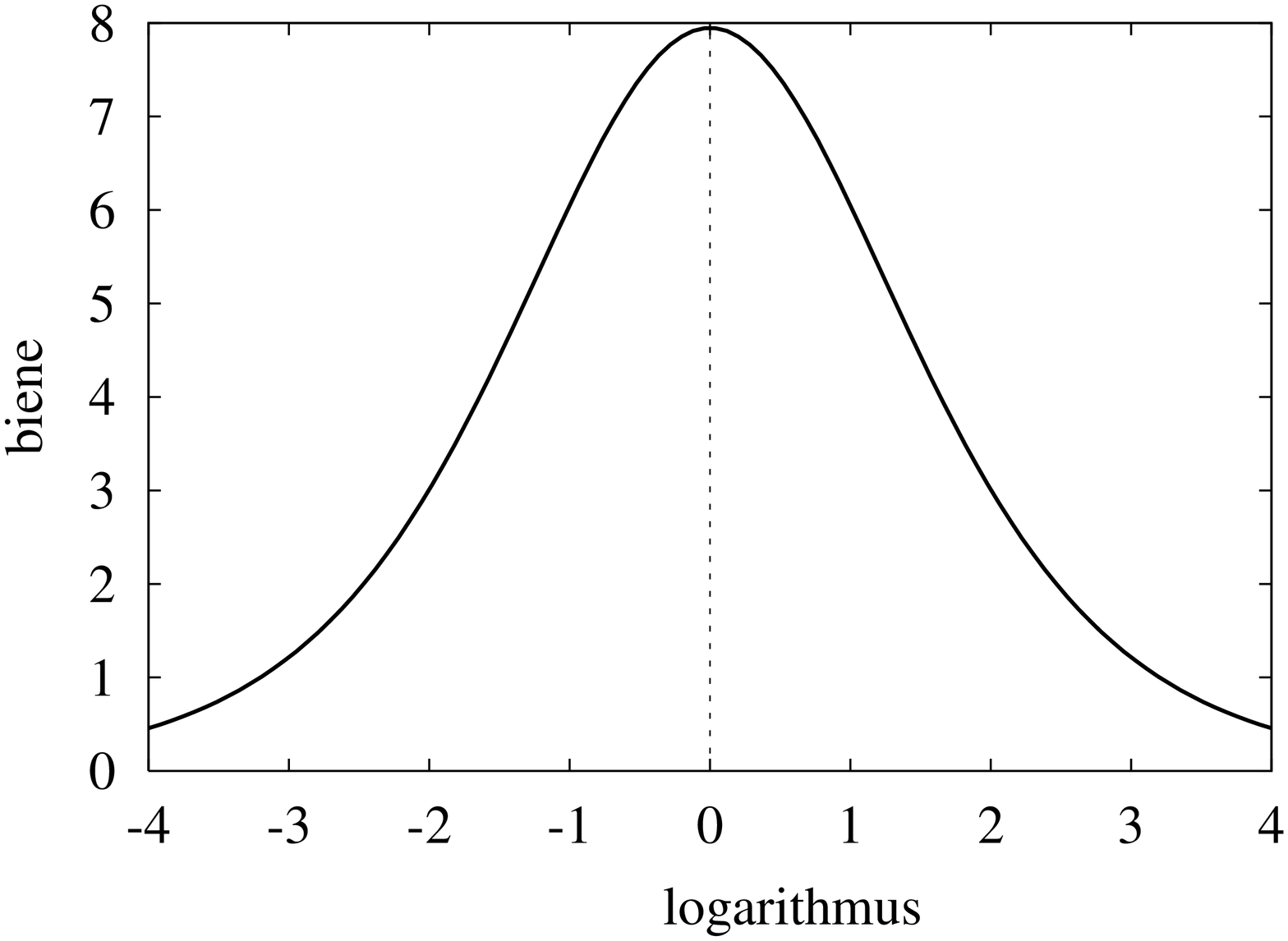}
\caption{\small{Colour zero modes $\lambda_a$}}
\label{fig:o5_a_ZM}
}
\parbox{6cm}{
\psfrag{logarithmus}{$\ln \left(\kappa\right)$}
\psfrag{rehlein}{$\| \widehat{\psi}^{(\alpha)}\|\big/\rpeak$}
\includegraphics[scale=0.25,angle=270]{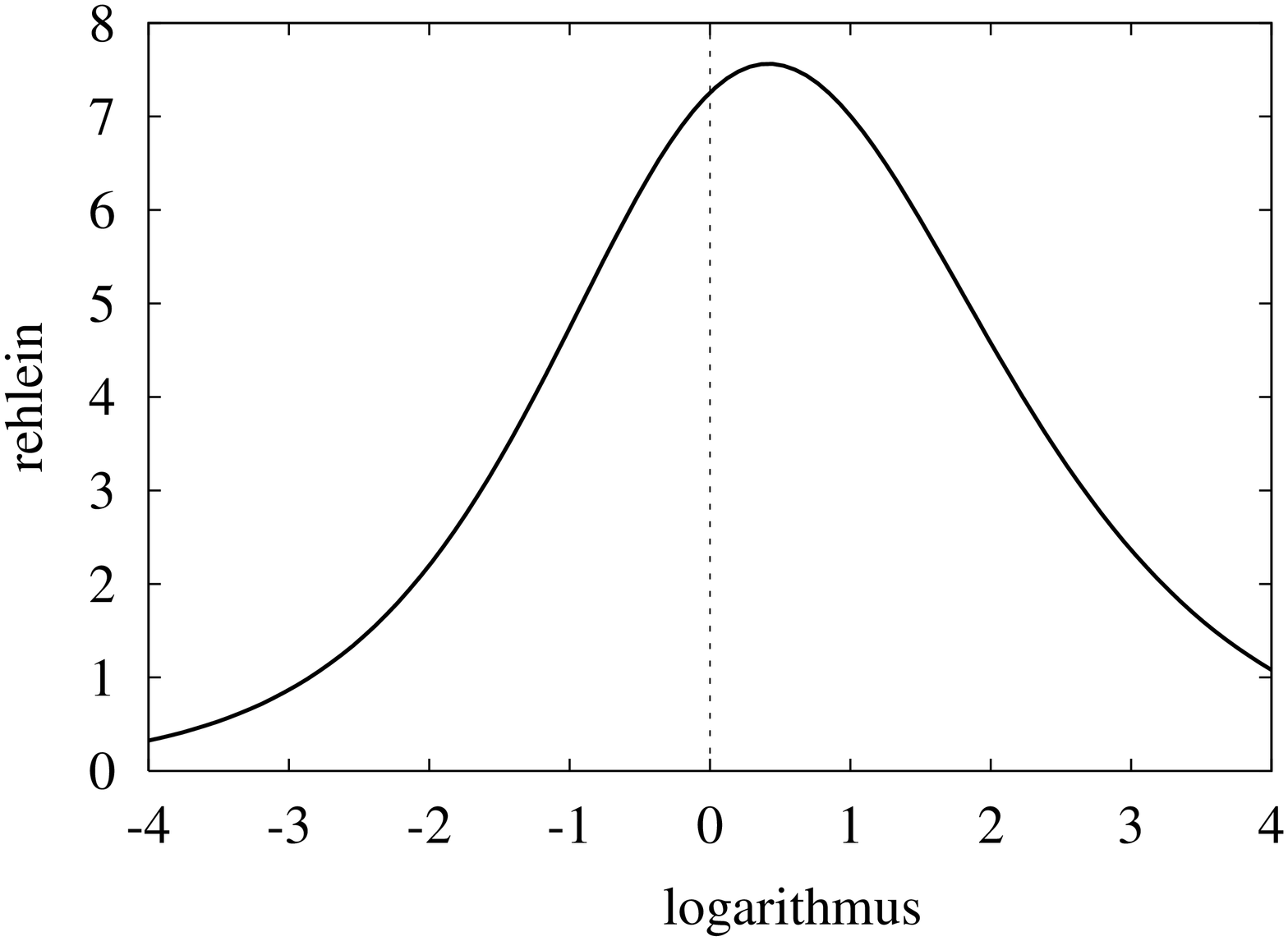}
\caption{\small{Colour zero modes $\lambda_{\alpha}$}}
\label{fig:o5_alpha_ZM}
}
\hspace{2cm}
\parbox{6cm}{
\psfrag{logarithmus}{$\ln \left(\kappa\right)$}
\psfrag{blume}{$\| \widehat{\psi}^{(z)}\|$}
\includegraphics[scale=0.25,angle=270]{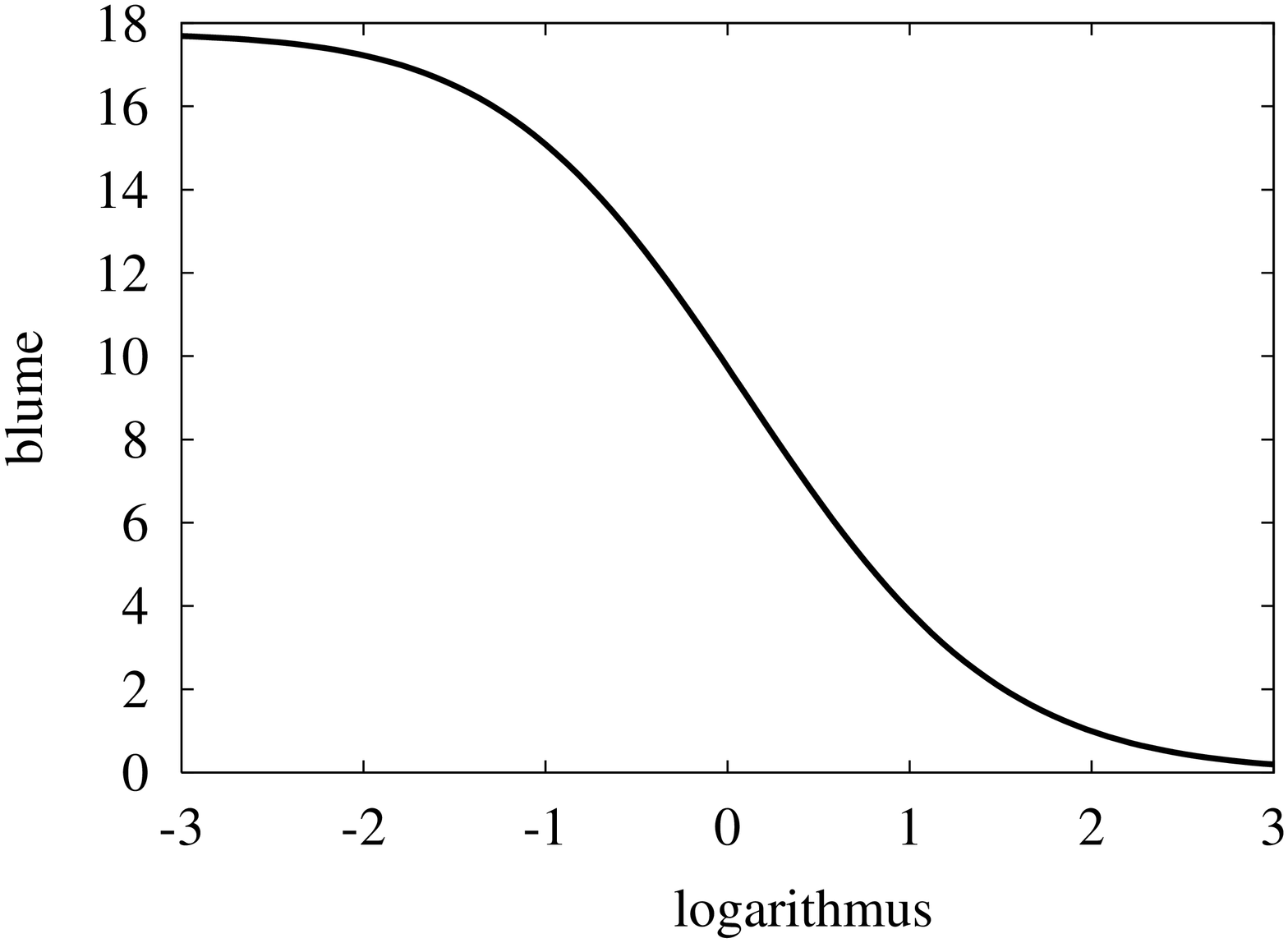}
\caption{\small{Translation zero modes}}
\label{fig:o5_trans_ZM}
}
\end{center}
\end{figure}

In Figs. \ref{fig:o5_dil_ZM} - \ref{fig:o5_trans_ZM}, the normalisation integrals of the four different types of zero modes are displayed versus $\ln\left(\kappa\right)$. 
It is apparent at first sight that except for the colour zero modes $\psi^{(a)}$ shown in \Fref{fig:o5_a_ZM}, the individual normalisation integrals are \emph{not manifestly symmetric} functions under instanton size inversion $\kappa\to1/\kappa$ i.e. $\ln\left(\kappa\right)\leftrightarrow -\ln\left(\kappa\right)$. Yet we recall that   
$\kappa\| \widehat{\psi}^{(\rho)}(\kappa)\|$ is symmetric as well. 
However, after evaluating the total zero mode contribution $\kappa^5\,\widehat{J}(\kappa)$ in \Eref{allzms}, along with the non-zero mode correction factor $1/\kappa$ via \Eref{dapprox}, we find the following, very promising results
depicted in \Fref{fig:o5_total_ZM_centred}:
\begin{figure}[h]
\parbox{8cm}{
\psfrag{logarithmus}{$\ln \left(\kappa/1.19\right)$}
\psfrag{total4}{$\kappa^4\,\widehat{J}(\kappa)\Big /\rpeak^7$}
\includegraphics[width=8cm,height=8cm,angle=270]{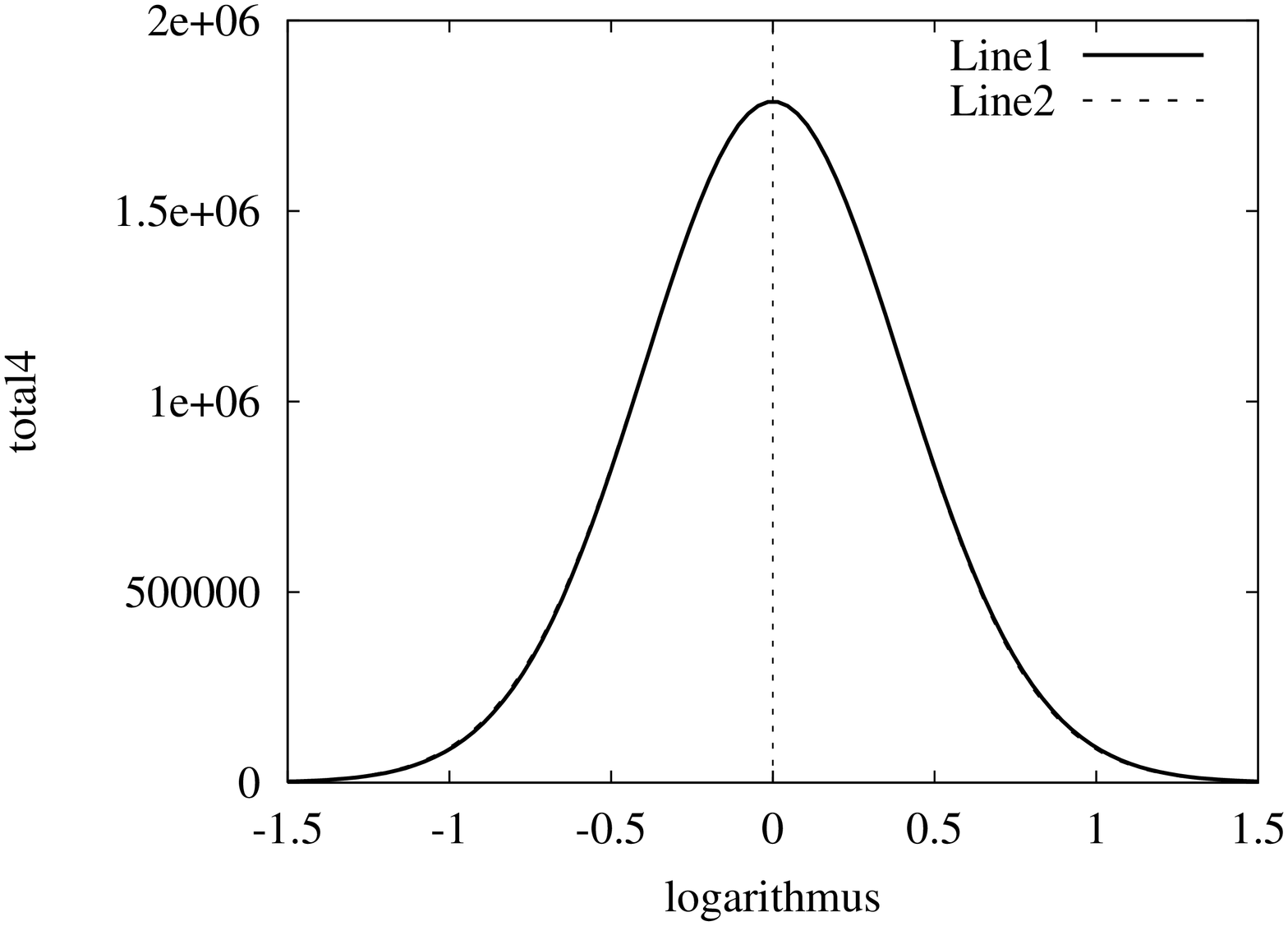}}
\hfill\parbox{8cm}{
\psfrag{logarithmus}{$\ln \left(\kappa/1.19\right)$}
\psfrag{total4}{$\kappa^4\,\widehat{J}(\kappa)\Big/\rpeak^7$}
\includegraphics[width=8cm,height=8cm,angle=270]{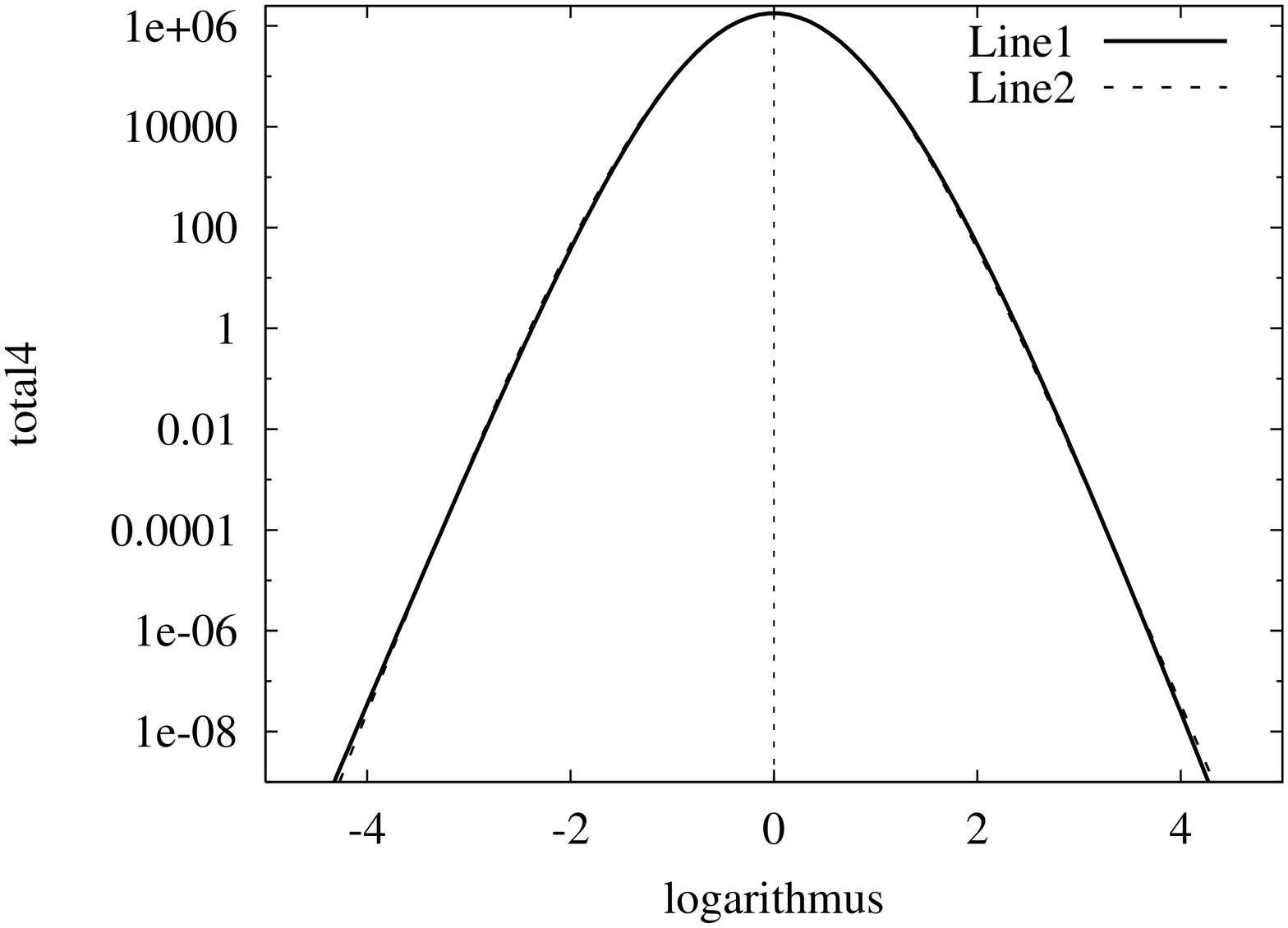}
}
\caption{\label{fig:o5_total_ZM_centred}(Left): Total zero mode contribution $\kappa^5\,\widehat{J}(\kappa)\big/\rpeak^7$ times $1/\kappa$ on the sphere, as relevant for the instanton size distribution $d(\rho\muren,\alpha_s)$ in \Eref{dapprox}. As abscissa, we use $\ln \left(\kappa/1.19\right)=\ln\left(\rho/(1.19\,\rpeak)\right)$ to make small deviations from a perfect size inversion symmetry for $\kappa/1.19\leftrightarrow 1.19/\kappa$ self-evident. Line1 and Line2 display the results versus $\kappa/1.19$ and $1.19/\kappa$, respectively. (Right): Same plot, but using a logarithmic ordinate, to demonstrate the high degree of inversion symmetry over fourteen orders of magnitude, with Line1 and Line2 being visually indistinguishable!}
\end{figure}
\begin{itemize}
\item Unlike conventional instanton perturbation theory, larger sized instantons are strongly suppressed in qualitative accordance with the lattice data (cf. \Fref{isize}) and general expectations.
\item Despite the fact that some zero mode norms are asymmetric under
$\kappa \leftrightarrow 1/ \kappa$, the \emph{total} zero mode contribution $\kappa^5\,\widehat{J}(\kappa)$ times the $1/\kappa$ non-zero mode correction leads nonetheless to an instanton size distribution~(\ref{dapprox}) that appears strikingly symmetric under this instanton size inversion after a slight rescaling 
\beqn\label{rescaling}
\kappa \rightarrow 1.19\kappa,
\eeqn
\item \Fref{fig:o5_total_ZM_centred} (right) illustrates impressively that the size distribution on the sphere and the inverted distribution are almost
indistinguishable over fourteen orders of magnitude!
\item The slight rescaling factor~(\ref{rescaling}) of $\mathcal{O}(1)$ might well find an explanation as a scheme-dependence effect within renormalisation-group considerations as in Sect.~\ref{sbeyond}
\end{itemize}
Let us discuss next some important \emph{analytical} properties and limits of the total zero mode part $\kappa^5\,\widehat{J}(\kappa)\Big/\rpeak^7$. Since the latter depends only on $\kappa$, the following two interesting interpretations of the limit $\kappa\to 0$ are identical with respect to our results: 
\begin{eqnarray*}
\kappa \rightarrow 0:\quad \left\{  \begin{array}{l}
\textrm{The radius $\rhop$ of the sphere tends to infinity for fixed instanton size $\rho$.}\\ 
\textrm{The instanton size $\rho$ tends to zero for fixed radius $\rhop$ of the sphere.}
\end{array}\right.
\end{eqnarray*}

\paragraph*{Limit of small instantons on the sphere:} 
\beqn
\kappa \rightarrow 0\;\Leftrightarrow \;\rhop \rightarrow \infty,\ \rho\ \textrm{fixed} \;\Leftrightarrow \;\rho \rightarrow 0,\ \rpeak\ \textrm{fixed to its physical value}
\eeqn
For $\rpeak\to \infty,\ \rho\ \textrm{fixed}$, we recover the same values for the zero mode normalisations as in flat Euclidean space (cf. \Eref{eq:zm_contribution}), apart from an additional factor of $2$ for every zero mode normalisation. It comes from the conformal factor of the stereographic projection:
\beqn
\widehat{\psi}_a \widehat{\psi}^a = \frac{4\,\rhop^4}{\left(\rhop^2+x^2\right)^2} \psi\umu \psi\omu \stackrel{\rhop\rightarrow \infty}=
4\,\psi\umu \psi\omu
\eeqn
The zero mode part $\rho^5\,\widehat{J}(\rho)$  rises as $\mathcal{O}\left(\rho^{12}\right)$ for small instantons in perfect agreement with instanton perturbation theory in flat Euclidean space~(\ref{eq:zm_contribution}).
As it should be in this regime of conventional instanton perturbation theory, any dependence of the size distribution on the new instanton scale $\rpeak$ drops out in this limit.

\paragraph*{Limit of large instantons on the sphere: }
\begin{align}
\kappa \rightarrow \infty \;\Leftrightarrow \;\rhop \rightarrow 0,\ \rho\ \textrm{fixed} \;\Leftrightarrow \; \rho \rightarrow \infty,\ \rpeak\ \textrm{fixed to its physical value}
\end{align}
We find that $\rho^5\,\widehat{J}(\rho)$  decreases like
$\mathcal{O}\left(1/ \rho^{12}\right)$, i.e. it decreases with the the same power as in the previous case, corresponding to asymptotic instanton size inversion symmetry!
\paragraph*{Limit for $\kappa = 1$:} This corresponds to the case of $\rhop=\rho$, 
which was considered in Refs.~\cite{Jackiw:1976dw,Ore:1976gb}. 
The expansion of $\rho^5\,\widehat{J}(\rho)$ about $\kappa=1$
reproduces these well-known results, as expected.

\section{Conformal Inversion and the Chirality-flip Ratio\label{sflip}}

It is clearly desirable to have another independent and preferably direct test of conformal inversion symmetry besides the instanton size distribution (cf. Fig.~\ref{isize}). In this section we shall briefly include an interesting such possibility from ongoing work~\cite{zs:2008}. 

If instantons are the dominant source of non-perturbative interactions in the QCD vacuum, then one should be able to observe that light quarks (zero modes) flip their chirality each time they cross the field of an instanton. Therefore, in Refs.~\cite{Faccioli:2002xf,Faccioli:2003qz}, the following chirality-flip ratio $R^\textrm{NS}$ was introduced  as function of Euclidean time t,
\begin{align}
R^\textrm{NS}(t)\equiv \frac{A^\textrm{NS}_\textrm{flip}(t)}{A^\textrm{NS}_\textrm{non-flip}(t)}= \frac{\Pi_\pi(t) - \Pi_\delta(t)}{\Pi_\pi(t) + \Pi_\delta(t)},
\label{rflip}
\end{align} 
in terms of the flavour non-singlet (NS) pseudo-scalar and scalar two-point correlators, 
\begin{align}
\begin{split}
\Pi_\pi(t)&=\langle 0\mid J_\pi(t)\,J_\pi^\dagger(0)\mid 0\rangle,\qquad J_\pi(x)=\overline{u}(x)i\gamma_5 d(x),\\
\Pi_\delta(t)&=\langle 0\mid J_\delta(t)\,J_\delta^\dagger(0)\mid 0\rangle,\qquad J_\delta(x)=\overline{u}(x)\,d(x).
\end{split}
\end{align}
Notice that the ratio $R^\textrm{NS}(t)$ must vanish as $t\to 0$ (no chirality flips), and must approach 1 as  $t\to \infty$ (infinitely many chirality flips).

\begin{figure}[ht]
\begin{center}
\includegraphics[width=10cm, angle=-90]{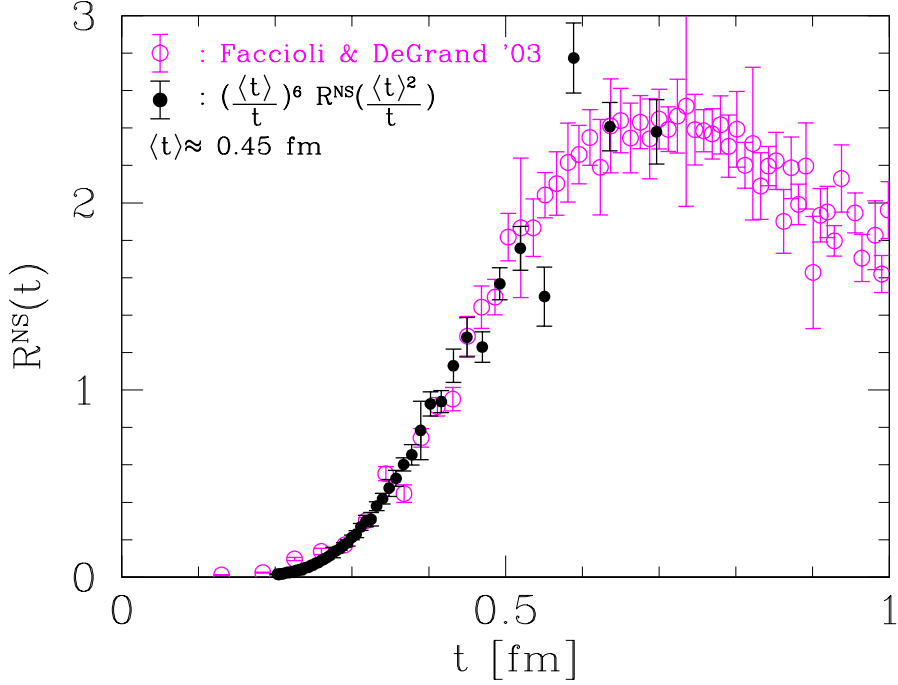}
\end{center}
\caption[dum]{\label{rflip-data} Alternative test of conformal (space-) time inversion symmetry for $t\to \langle t\rangle^2/t$, according to Eqs.~(\ref{rflip}, \ref{flipansatz}) for the chirality-flip correlator $R^\textrm{NS}(t)$, using directly (quenched) lattice data from Ref.~\cite{Faccioli:2003qz}. The abscissa denotes Euclidean time in fermi. The agreement of the direct and overlaid data provides strong evidence that conformal (space-)time inversion holds also in the fermionic sector.}
\end{figure}

These characteristic predictions for the chirality-flip ratio have been checked on the lattice in Ref.~\cite{Faccioli:2003qz} (within the quenched approximation). The results displayed in Fig.~\ref{rflip-data}, indeed provide impressive evidence for instanton dominance, as was discussed in detail in Ref.~\cite{Faccioli:2003qz}.
 
After these prerequisites, let us turn next to the important question whether these lattice data for $R^\textrm{NS}$, might show evidence for conformal space-time inversion symmetry at work? We expect the following general time inversion law for a scalar observable~\cite{diFrancesco:1997} with  conformal scaling dimension~$\Delta$
\begin{align}  
R^\textrm{NS}\left(t\right ) = \left(\frac{t}{\langle t\rangle}\right)^{2\,\Delta}\,R^\textrm{NS}\left(\frac{\langle t\rangle^2}{t}\right). 
\label{flipansatz}
\end{align}
$\Delta$ can be immediately fixed from the two general requirements that $R^\textrm{NS}(t)$ approaches a nonvanishing constant ($=1$ ) for large t, while -- within an instanton framework -- it vanishes $\propto t^6$ for $t\to 0$, due to the known behaviour of (the square of) the non-zeromode propagator in the denominator of the chirality-flip ratio~(\ref{rflip}). We then immediately infer from the ansatz~(\ref{flipansatz}),  
\begin{align}
\Delta=3.
\end{align}
In Fig.~\ref{rflip-data}, the prediction $\left(\,t/\langle t\rangle\,\right)^6\,R^\textrm{NS}\left(\langle t\rangle^2/t\right)$ from conformal inversion~(\ref{flipansatz}) with $\Delta=3$, and invoking the lattice data at {\em large $t$ values} ($t\,\gwig\, \langle t\rangle\,\approx 0.45$ fm), is overlaid on the lattice data for {\em small $t\,\lwig\, \langle t\rangle$}.

Apparently, the agreement is virtually perfect and thus supports strongly the validity of conformal space-time inversion. Note that this is not only an independent test using directly lattice data, but also it probes for the first time conformal inversion properties in the fermionic sector.   
    
\section{Implications beyond Instantons\label{beyond}\label{sbeyond}}

In this section, we shall argue that the advocated
$\rho\to \rpeak^2/\rho$ inversion symmetry of the instanton size distribution will affect the form of $\alpha_s$, and thus has implications for QCD in general. 

With the size distribution being a (lattice) observable, it must be renormalisation-group invariant. Indeed, for the perturbative expression \Eref{density}, one finds at the two-loop level, with $\alpha_s(\muren)=\alpha^\textrm{2-loop}_s(\muren)$,
\begin{align}
\begin{split}
\exp{\left(-\frac{2\pi}{\alpha_s(\muren)}\right)}\,\left(\rho\muren\right)^{\beta_0+\beta_1\frac{\alpha_s(\muren)}{4\pi}} &=\exp{\left(-\frac{2\pi}{\alpha_s(\frac{1}{\rho})}\right)}\,\left( 1+\mathcal{O}(\alpha_s^2(\muren)\ln(\rho\muren)^2\right);\\
\left(\frac{2\pi}{\alpha_s(\muren)}\right)^{2N_c}\left(\rho\muren\right)^{-4 N_c\,\beta_0\,\frac{\alpha_s(\muren)}{4\pi}}   &=\left(\frac{2\pi}{\alpha_s(\frac{1}{\rho})}\right)^{2N_c}\,\left( 1+\mathcal{O}(\alpha_s^2(\muren)\ln(\rho\muren)^2\right).
\end{split}
\end{align}
Hence, at two-loop level, the perturbative instanton size distribution~(\ref{density}) may be rewritten in manifestly renormalisation group invariant form,
\begin{align}
d\left(\rho\muren,\,\alpha_s(\mu_r)\right) = d\left(1,\alpha_s(\frac{1}{\rho})\right)
=C\left(\frac{2\pi}{\alpha_s(\frac{1}{\rho})}\right)^{2N_c}\,\exp\left(-\frac{2\pi}{\alpha_s(\frac{1}{\rho})}\right) = \rho^5\,\frac{\ad n^{(I)}}{\ad^4 z \,\ad\rho}  
\label{rginv}
\end{align}
with {\em all} $\rho$ dependence now residing in the running coupling $\alpha_s(\frac{1}{\rho})$. Next, we follow Ref.~\cite{Ringwald:1999ze} and define a (non-perturbative) ``instanton scheme'' for the running coupling,  
\begin{align}
\alpha^I_s(\frac{1}{\rho}) = \alpha^{\overline{MS}}_s\,(\frac{s^I}{\rho})\equiv \alpha_s\,(\frac{s^I}{\rho}),
\end{align}
by the requirement that the perturbative expression~(\ref{density}) of the density, involving the two scheme-independent $\beta$-function coefficients $\beta_0,\beta_1$, remains valid for {\em all values} of $\alpha^I_s$. Surprisingly, the form of $\alpha^{\overline{MS}}_s(\frac{s^I}{\rho})$, implicitly defined by this prescription and directly extracted from a comparison with the UKQCD data~\cite{Smith:1998wt,Ringwald:1999ze} for the instanton size distribution (cf. \Fref{isize}), showed a (confining)
Cornell form $\alpha_s \approx \frac{3}{4}\, \sigma\,\rho^2+\ldots$ for
$\rho\gwig\rpeak$ with string tension $\sqrt{\sigma}\approx 472$ MeV, while beautifully agreeing with the 3-loop perturbative form of $\alpha^{\overline{MS}}_s$ for $\rho\lwig \rpeak$. In addition, the resulting scale factor~\cite{Ringwald:1999ze}, $s^I = \Lambda_{\overline{MS}}/\Lambda_I\approx 1.18=\mathcal{O}(1)$, puts the "instanton scheme" very close to the $\mathrm{\overline{MS}}$ scheme in the perturbative regime!

After these relevant prerequisites, let us combine the requirements of renormalisation-group invariance~(\ref{rginv}) and conformal space-time invariance of the instanton size distribution $d\left(\rho\muren,\,\alpha_s(\mu_r)\right)$,
\begin{align}
d\left(1,\,\alpha^I_s(\frac{1}{\rho})\right) = \left( \frac{\rpeak}{\rho}\right)^{2\,\Delta}\,d\left(1,\,\alpha^I_s(\frac{\rho}{\rpeak^2})\right).
\label{dinv}
\end{align}
In the present more general context, and in analogy to \Eref{flipansatz} of the preceeding section~\ref{sflip}, we have allowed for a (small) non-vanishing conformal scaling dimension $\Delta$ that may balance the remaining uncertainty concerning the infrared behaviour of the non-zero mode part $Q(\gamma)$ (cf. \Eref{nzm}). With the form~(\ref{rginv}) in the "instanton scheme", \Eref{dinv} implies the relation,
\begin{align}    
\left(\frac{2\pi}{\alpha^I_s(\frac{1}{\rho})}\right)^{2
N_c}\,\exp(-\frac{2\pi}{\alpha^I_s(\frac{1}{\rho})})=\left( \frac{\rpeak}{\rho}\right)^{2\,\Delta}\,\left(\frac{2\pi}{\alpha^I_s(\frac{\rho}{\rpeak^2})}\right)^{2
N_c}\,\exp(-\frac{2\pi}{\alpha^I_s(\frac{\rho}{\rpeak^2})}),
\label{sym}
\end{align}
the solution of which relates the running coupling $\alpha_s$ in the asymptotically free ($\rho \lwig \rpeak$) and confining regimes ($\rho\gwig\rpeak$)! 
 
The solution for $\alpha^I_s(1/\rho)$ in terms of $\alpha^I_s(\rho/\rpeak^2)$  takes a simple and intriguing form,
\begin{align}
\frac{\pi}{\alpha^I_s\left(\frac{1}{\rho}\right)\,N_c}=-W\left( -\frac{\pi}{\alpha^I_s(\frac{\rho}{\rpeak^2})\,N_c} \,\exp\left[-\frac{\pi} {\alpha^I_s(\frac{\rho}{\rpeak^2})\,N_c}\right] \left(\frac{\rpeak}{\rho}\right)^{\frac{\Delta}{N_c}} \right),
\label{sol}
\end{align}
involving the Lambert W function~\cite{Corless:1997},   
\begin{eqnarray}
W(x)\,e^{W(x)}=x;&
\mathrm{with\ two\ real\ branches\ } W_0(x)\mathrm{\ and\ } W_{-1}(x) \mathrm{\ for\ } -1/e\le x <0,\label{w1}\\
&\mathrm{satisfying\ } W_0(-1/e)=W_{-1}(-1/e)= -1.\label{w2}
\end{eqnarray}
Note that 
\begin{align}
\mathrm{for\ } x\to 0^-:\ \left\{\begin{array}{lcl}
W_0(x)&\sim & x-x^2+\mathcal{O}(x^3)\to 0,\  (\mathrm{principal\ branch\ } )\\ 
W_{-1}(x)&\sim& \ln(-x)-\ln(-\ln(-x))\ldots \to -\infty.
\end{array}\right. 
\label{asym}
\end{align}
The solution~(\ref{sol}) exhibits a number of remarkable features to which we turn next.

First of all, let us insert the leading, asymptotically free expression of the running coupling for large $\rho>\rpeak$,
\begin{align}
\alpha^I_s(\frac{\rho}{\rpeak^2})\approx
\frac{2\pi}{\beta_0}\,\frac{1}{\log(\frac{\rho}{\rpeak^2\,\Lambda_I})}+\ldots,\end{align}
into the r.h.s of the inversion-symmetry solution~(\ref{sol}).
With the assignment~\cite{Schrempp:2001ir}
\begin{align}
\Delta=\frac{N_c}{6}\ \stackrel{\mathrm{SU(3)}}{=}\frac{1}{2},
\label{delta}
\end{align}
of the conformal scaling dimension~(\ref{dinv}) and the asymptotics~(\ref{asym}) of $W_0(x)$ for $x\to 0^-$, we find once more a Cornell form of the running coupling for large $\rho$   
\begin{equation}
\frac{\alpha^I_s(\frac{1}{\rho})\,N_c}{\pi} \approx\frac{6}{11}\,\frac{1}{[\rpeak \Lambda_I]^{11/6}\,\,\ln\left(\frac{\rho}{\rpeak^2\Lambda_I} \right)}\,\left(\frac{\rho}{\rpeak}\right)^2 - 1 + o\left(\left(\frac{\rpeak}{\rho}\right)^2\right),  
\label{cornell1}
\end{equation}
signalling confinement. In Ref.~\cite{Ringwald:1999ze}, numerical agreement with a Cornell form for $\alpha^I_s$ was observed after solving \Eref{rginv} for $\alpha^I_s(1/\rho)$ in terms of the UKQCD {\em lattice data}~\cite{Smith:1998wt,Ringwald:1999ze} for $\rho^5\,\frac{\ad n^{(I)}}{\ad^4 z \,\ad\rho}$. Here, we obtained the same result in analytical form, only from conformal inversion symmetry~(\ref{sol}) and the known short-distance behaviour of $\alpha_s$. 

The non-vanishing conformal scaling dimension~(\ref{delta}) implies a slight deviation from the simplest assumption~(\ref{nzm}) of a uniform correction to the $\rho$ dependence via the non-zero mode factor $Q(\gamma)$ in the size distribution. At short distances (small $\rho$), the inversion law~(\ref{dinv}) together with $\Delta$ from \Eref{delta}, corresponds to the perturbative correction $Q \sim (\muren\,\rho)^{-N_c/3}$  as in \Eref{nzm}. However, at long distances (large $\rho$), the $\rho$-dependence arises {\em only} from the calculable zero-mode part $J(\gamma)$, like in supersymmetric Yang-Mills theory~\cite{Novikov:1983uc}.

Moreover, we note that with the above scaling dimension~(\ref{delta}), $\Delta \propto N_c$, the solution~(\ref{sol}) only depends on the 't Hooft coupling~\cite{'tHooft:1973jz}, $g^2_s\,N_c \propto \alpha_s\,N_c$, such that it remains {\em unchanged} in the large $N_c$ limit. 
   
For $\rho=\rpeak$, \Eref{sol} may be solved for it's only unknown $\frac{\pi}{\alpha^I_s(1/\rpeak)\,N_c}$, along with the matching condition 
\begin{align}
u\equiv-W_0(-u\,e^{-u})=-W_{-1}(-u\,e^{-u}); \mathrm{\ where\ }u=\frac{\pi}{\alpha^I_s\left(\frac{1}{\rpeak}\right)\,N_c}.
\end{align}
One finds the unique solution (due to \Eref{w2}),
\begin{align}
\frac{\alpha^I_s\left(\frac{1}{\rpeak}\right)\,N_c}{\pi} = \frac{\alpha^{\overline{MS}}_s\,\left(\frac{s^I}{\rpeak}\right)\,N_c}{\pi} = 1,
\label{peaknorm}
\end{align} 
that indeed matches the peak position of the instanton size distribution~(\ref{rginv}), i.e. $\frac{\ad }{\ad\,\alpha^I_s}\, d(1,\alpha^I_s)=~0$, as function of $\alpha^I_s(\frac{1}{\rho})$.
 
Last not least, one may write down an oversimplified
but {\it exact} closed solution of \Eref{sol}, 
\begin{equation}
\alpha^I_s(\frac{1}{\rho})=\frac{2\,\pi}{\beta_0}\,\frac{(1-\left( 
\frac{\rho}{\rpeak}\right)^2)}{\ln(\frac{\rpeak}{\rho})},\qquad \Lambda\approx\frac{1}{\rpeak},
\label{nesterenko}
\end{equation}
which follows upon requiring in addition to \Eref{sol} the inversion symmetry,  
\begin{equation}
\left(\frac{\rho}{\rpeak^2}\right)\,\alpha^I_s\left(\frac{\rho}{\rpeak^2}\right)=\left(\frac{1}{\rho}\right)\, \alpha_s\left(\frac{1}{\rho}\right).  
\end{equation}
Despite it's simplistic form, \Eref{nesterenko}
has {\it no Landau pole} for $\rho\to\rpeak$, exhibits the correct  asymptotic freedom behaviour for
$\rho\Rightarrow 0$, as well as a Cornell form (\ref{cornell1}) for large $\rho$.
The peak normalization condition~(\ref{peaknorm}) only holds approximately, $\alpha^I_s(1/\rpeak)\,N_c/\pi = 12/11\approx 1$, but can be satisfied with a slightly more complex limiting process. 
Amazingly, this (1-loop) form (\ref{nesterenko}) of $\alpha_s$ exists
already in the literature\cite{Nesterenko:2003xb}, but originated from an entirely different reasoning.  It appeared as the appropriate (1-loop) running coupling without a
Landau pole in a renormalisation-group improved variant of
Shirkov's ``analytic perturbation theory''~\cite{Shirkov:2006gv}.

\section{Conclusions}
In the present investigation, we have studied the appealing possibility that the strong suppression of large-size QCD instantons -- as evident from lattice data -- is due to a surviving conformal space-time inversion symmetry. 

We started from the known fact that the classical instanton sector is conformally invariant and notably also invariant under conformal space-time inversion $x\umu\to x\umu\oprime =\frac{b^2}{x^2}\,x\umu$. Since the latter acts non-infinitesimally like a discrete symmetry transformation, it is not a generator of the conformal group. Yet all conformal generators can be composed of an even number of inversions and generators of the Poincar\'e subgroup.  This inversion symmetry is both suggested from the striking invariance of high-quality lattice data for the instanton size distribution under inversion of the instanton size $\rho \to \langle\rho\rangle^2/\rho$ (cf. \Fref{isize}) and from the known validity of space-time inversion symmetry in the classical instanton sector. 

Our theoretical line of attack in this paper was restricted to a detailed study of the {\em zero-mode part} of the instanton size distribution, which we have argued to constitute the "dominating" source of the $\rho$-dependence. In this context, it is most encouraging that the instanton size distribution of {\em supersymmetric} Yang-Mills theories is known to be entirely given in terms of zero-modes~\cite{Novikov:1983uc}.

A main theoretical step consisted in performing a conformal stereographic projection of the instanton calculus in flat Euclidean space to the 4-dimensional surface of a 5-dimensional sphere. This way,
we have achieved several benefits at once.  
\begin{itemize}
\item All zero-mode normalisation integrals on the sphere remained finite under space-time inversion, since the sphere represents a compact, curved geometry. 
\item The identification of the sphere radius $b\equiv\rpeak$ provided a natural way of introducing the crucial physical scale $\rpeak$ into the
instanton calculus. It acts as the conformal inversion radius.
\item On the sphere surface, the normalization integrals all turned out to be {\em invariant} under space-time inversion due to a "fortunate cooperation" of the scale factors associated with both the conformal stereographic projection and space-time inversion.  
\item While the zero-mode normalizations $\| \widehat{\psi}^{(a)}\|,\ a=1,2,3$ (colour) and $\rho/\rpeak\,\| \widehat{\psi}^{(\rho)}\|$ (dilatation) are indeed manifestly symmetric under $\rho \to\rpeak^2/\rho$, the remaining zero-mode norms are not. However the product of all of them, as entering the instanton size distribution, is symmetric to an impressively high degree. Altogether, the resulting shape due to the product of zero-modes is in good qualitative agreement with the lattice data (cf. \Fref{isize}, \Fref{fig:o5_total_ZM_centred}), strongly suppressing large-size instantons!  
\item The present formulation on the sphere allowed to study various limits of theoretical interest, which underligned the consistency of the present approach: notably, the limit $\rho/\rpeak\to 0$ may either be viewed as a "flat-space" limit (sphere radius $\rpeak\to\infty$) with instanton size $\rho$ kept fixed, or as the small instanton limit ($\rho\to 0$) with the sphere radius $\rpeak$ kept fixed. Irrespectively, for $\rho/\rpeak\to 0$, we recover the familiar results of instanton perturbation theory in flat 4-dimensional Euclidean space. 
\end{itemize} 
As important, independent and direct further support for conformal inversion symmetry at work, we presented the striking evidence from a lattice simulation of the chirality-flip ratio $R^\textrm{NS}$ in the QCD vacuum as function of Euclidean time~\cite{Faccioli:2002xf,Faccioli:2003qz}.

Finally, we explored some striking consequences of conformal space-time inversion symmetry beyond instantons, i.e. for QCD in general. It implied a general relation between the running coupling at short and long distances. From the familiar input of asymptotic freedom at short distances, we found a Cornell form $\alpha^I_s(1/\rho)\propto \sigma\, \rho^2$ at long distance, signalling confinement.

\section*{Acknowledgements}
One of us (D.\,K.) is grateful to Gerard 't Hooft for a very helpful communication. F.\,S. wishes to thank Mikhail Shifman for interesting discussions and suggestions and Tom DeGrand for making his lattice data available. D.\,K. is grateful for the hospitality extended to her at DESY, where this paper was completed. Moreover, she acknowledges financial support by the FWF Project P20017. We are grateful to Bryan Zald\'ivar Montero (Havana/Cuba) for investigating some related questions during his six months visit at DESY with a fellowship of the High Energy Physics Latinamerican-European Network (HELEN).   


\section*{Appendix A: Conformal Transformations}\label{app: CT}
In this paper we consider active conformal transformations~\cite{Fulton:1976,deAlfaro:1973,diFrancesco:1997} throughout,
\begin{equation}\label{defCT}
g_{\mu\nu}(x) \frac{\partial x^{\mu}}
{\partial x^{\prime\kappa}}
\frac{\partial x^{\nu}}{\partial x^{\prime\lambda}}= \sigma(x)\; g_{\kappa \lambda}(x\oprime)=g\oprime_{\kappa\lambda}(x\oprime),
\end{equation}
where $\sigma(x)$ is called \textit{conformal} or \textit{scale factor}.

The conformal inversion, 
\begin{equation}\label{eq:def-inv}
x\umu \rightarrow x\umu\oprime=\frac{b^2}{x^2}x\umu,
\end{equation}  
is the relevant transformation for our approach. $b$ is called the radius of inversion. The scale factor is given as
$\sigma(x)_{\textrm{inv}}=x^4/b^4$.
The transformation law for a covariant vector field under an active transformation is given by
\begin{align}\label{eq:coVF_law}
A\umu\oprime(x\oprime)&=\frac{\partial x\onu}{\partial x^{\prime\mu}}A\unu(x)=\sqrt{\sigma(x)}I\umu^{\phantom{i}\nu}(x)A\unu(x),
\end{align}
whereas the corresponding contravariant vector field has to transform as~\cite{Fulton:1976},
\begin{align}\label{eq:contraVF_law}
A^{\prime\mu}(x\oprime)&=\sigma(x)\frac{\partial x^{\prime\mu}}{\partial x\onu}A\onu(x)=\sqrt{\sigma(x)}I\omu_{\phantom{i}\nu}(x) A\onu(x),
\end{align}
where the scale factor $\sigma(x)$ appears when pulling up the index with the metric  
tensor\footnote{In the
case of a passive transformation this factor would not be included.}.
The tensor~\cite{Shintani:1982vp} 
\begin{align}
I^{\phantom\mu\nu}\umu(x)=\frac{1}{\sqrt{\sigma(x)}}\frac{\partial x\onu}{\partial x^{\,\prime\mu}}.
\end{align}
satisfies in Euclidean space-time
\begin{align}
I\onu_{\phantom\nu\mu}(x)I\omu_{\phantom\mu\rho}(x)=\delta\onu_{\phantom\nu\rho}.
\end{align}
For a conformal space-time inversion it is given by
\begin{align}
I^{\textrm{inv}}_{\mu\nu}(x)=\delta_{\mu\nu} -\frac{2\,x\umu x\unu}{x^2}.  
\end{align} 
Furthermore the length of a vector is not invariant under
conformal transformations, in
fact it is stretched by the scale factor,
\begin{align}\label{eq: length_inv}
A^{\prime\mu}(x\oprime)A\umu\oprime(x\oprime)=\sigma(x)A\omu(x)A\umu(x).
\end{align}
The generalisation to second order tensors is straight forward. The volume element changes according to
\begin{align}
\ad^4 x\oprime =\frac{\ad^4 x }{\sigma^2(x)}.
\end{align}

\paragraph*{Transformation rules for stereographic projection}
\begin{align}\label{eq: stp trafo rules}
\begin{split}
\widehat{A}^a&=\sigma(x)_{\textrm{sp}}\frac{\partial r^a}{\partial x\omu}A\omu(x),\\
\widehat{A}_a&=\left(\frac{\partial}{\partial r^a}-\frac{1}{\rpeak^2}r_a(r\cdot\partial)\right)(x\omu)\,A\umu(x),\\
\widehat{A}_a(r)\widehat{A}^a(r)&=\spf\;A\umu(x)A\omu(x),\\
r_a\widehat{A}^a(r)&=0. 
\end{split}
\end{align}
The stereographic projection is done in such a way that this
constraint comes naturally with the transformation rules to ensure that the projected vector fields indeed stay on the sphere. 

\section*{Appendix B: Zero modes for $SU(3)$}
For completeness, we list here the twelve $SU(3)$ zero modes as derived by
Bernard~\cite{CB:qhe} and used in this paper. 

An $SU(3)$ instanton is obtained by embedding the $SU(2)$ instanton into the "upper-left-hand corner" of the fundamental representation of $SU(3)$. Thus, in singular gauge, an $SU(3)$ instanton takes the form
\begin{align}
A\umu^{(I)}(x)=\frac{1}{\sqrt{\pi\alpha_s}}\frac{\bar{\eta}_{a\mu\nu}x\onu}{x^2\left(\rho^2+x^2\right)}\frac{\lambda_a}{2}, 
\end{align}
where $\lambda_a$ are the first three Gell-Mann matrices and $\bar{\eta}_{a\mu\nu}$ are the 't Hooft coefficients~\cite{'tHooft:1976fv} for $SU(2)$. Latin indices $a$ and $b$ run from 1 to 3 and Greek indices $\mu$ and $\nu$ are space-time indices running from 1 to 4.

The zero modes are in background gauge with respect to the classical instanton field,
\begin{align}
D\umu^{\text{cl}}\psi\umu^{(i)}=\partial\umu \psi\umu^{(i)}-\ii\sqrt{4\pi\alpha_s} \left[A^{(I)}\umu,\psi^{(i)}\umu\right]=0.
\end{align}

\paragraph*{Dilatation zero mode:}
\begin{align}
\begin{split}
\psi^{(\rho)}\umu(x)
&=\frac{2}{\sqrt{\pi\alpha_s}}\frac{\rho \;\bar{\eta}_{a \mu\nu}\;x\onu}{(x^2+\rho^2)^2}\frac{\lambda_a}{2}
\end{split}
\end{align}

\paragraph*{Colour zero  modes for generators $\lambda_a;\ a=1,\,2,\,3$:}
\begin{align}\label{azm}
\begin{split}
\psi^{(a)}\umu(x)
&=\frac{1}{2\sqrt{\pi\alpha_s}}\frac{\rho^2}{(x^2+\rho^2)^2}\Big( 2x\umu \lambda^a -i \;\bar{\eta}_{b\mu\nu}x\onu\;
[\lambda_b,\lambda_a]\Big).
\end{split}
\end{align}

\paragraph*{Colour zero modes for generators $\lambda_{\alpha};\ \alpha=4,\,5,\,6,\,7$:}
\begin{align}\label{alphazm}
\begin{split}
\psi^{(\alpha)}\umu
&= \frac{1}{2\sqrt{\pi\alpha_s}}\frac{\rho^2}{(x^2)^{1/2}(x^2+\rho^2)^{3/2}}\big( x\omu \lambda_{\alpha}
-i\; \bar{\eta}_{b\mu\nu}\;x\onu [\lambda_b,\lambda_{\alpha}]\big)
\end{split}
\end{align}
Here $\lambda_{\alpha}$ denote the four Gell-Mann matrices $\lambda_4 \ldots\lambda_7$. 

\paragraph*{Translation zero modes:}
\begin{align}
\begin{split}
\psi^{\,(z\unu)}\umu(x)&=\frac{\partial A\umu^{(I)}(x-z)}{\partial z\onu} \Big|_{z=0}
+D^{\text{cl}}\umu(A^{(I)}\unu(x))
\\
&= -\partial\unu A^{(I)}\umu(x) + \partial\umu A^{(I)}\unu(x) +\ii \sqrt{4\pi\alpha_s} [ A^{(I)}\umu(x),A^{(I)}\unu(x)]
\\
&=-\frac{4}{\sqrt{\pi\alpha_s}}\frac{\rho^2}{\left(\rho^2+x^2\right)^2}
\left[ \frac{x\umu x^{\sigma}}{x^2}-\frac{1}{4}\delta\umu^{\sigma}\right]\bar{\eta}_{a\nu\sigma}\frac{\lambda_a}{2}
-\left(\mu \leftrightarrow \nu\right)
\\
&=F\umunu(x).
\end{split}
\end{align}

\bibliography{qhe} 
\end{document}